\documentclass[aoas,preprint, 11pt, dvipsnames, table, x11name]{imsart}

\usepackage{tikz-cd}
\usepackage[toc,page]{appendix}
\usepackage[font={small}, labelfont=bf]{caption}
\usepackage{subcaption}
\usepackage[utf8]{inputenc}
\newcommand{\E}{\mbox{E}}
\newcommand{\N}{\mbox{N}}
\usepackage{amsfonts}
\usepackage[T1]{fontenc}
\usepackage{comment}
\usepackage{xcolor}
\definecolor{darkgreen}{RGB}{0,69,0} 
\definecolor{navy}{RGB}{0,60,113} 
\usepackage{booktabs}
\usepackage{bbm}
\usepackage{array}
\newcolumntype{P}[1]{>{\centering\arraybackslash}p{#1}}
\newcolumntype{P}[1]{>{\centering\arraybackslash}p{#1}}

\usepackage{imakeidx}
\usepackage{tikz}
\usetikzlibrary{arrows, arrows.meta, calc, positioning}
\usetikzlibrary{positioning}
\newdimen\nodeDist{}
\usepackage{pgfplots}

\usepackage{graphicx}
\usepackage{amsmath,commath,amssymb,blkarray,bm,bbm}
\usepackage{bbm}
\usepackage{mathtools}

\renewcommand{\bm}[1]{\mathbf{#1}}
\usepackage{mathrsfs}
\usepackage{physics}
\usepackage[pagebackref]{hyperref}  
\usepackage{comment}
\usepackage[authoryear]{natbib}
\newcommand\independent{\protect\mathpalette{\protect\independenT}{\perp}}
\def\independenT#1#2{\mathrel{\rlap{$#1#2$}\mkern2mu{#1#2}}}

\newcommand{\ind}[1]{\mathbbm{1}({#1})}%
\hypersetup{
	colorlinks=true,
	linkcolor={blue!60!black},
	filecolor=magenta,      
	urlcolor={blue!60!black},
	citecolor={blue!60!black}
}
\RequirePackage{amsthm,amsmath,amsfonts,amssymb}
\RequirePackage[authoryear]{natbib}

\startlocaldefs

\theoremstyle{remark}

\endlocaldefs

\begin{document}
	
	\begin{frontmatter}
		\title{Do forecasts of bankruptcy cause bankruptcy? \\A machine learning sensitivity analysis.}
		\runtitle{Do Forecasts of Bankruptcies cause Bankruptcies?}
		
		\begin{aug}
			\author[A]{\fnms{Demetrios} \snm{Papakostas}\ead[label=e2,mark]{dpapakos@asu.edu}}
			\author[A]{\fnms{P. Richard} \snm{Hahn}\ead[label=e1,mark]{prhahn@asu.edu}},
			\author[B]{\fnms{Jared} \snm{Murray}\ead[label=e3,mark]{jared.murray@mccombs.utexas.edu}}
			\and
			\author[C]{\fnms{Frank} \snm{Zhou}\ead[label=e4,mark]{szho@wharton.upenn.edu}}
			\author[D]{\fnms{Joseph} \snm{Gerakos}\ead[label=e5,mark]{Joseph.J.Gerakos@tuck.dartmouth.edu}}
			
			\address[A]{School of Mathematical and Statistical Sciences,
				Arizona State University,
				\printead{e1,e2}}

			\address[B]{Department of Information, Risk and Operations Management,
				The University of Texas at Austin,
				\printead{e3}}

			\address[C]{The Wharton School,
				University of Pennsylvania,
				\printead{e4}}
			
			\address[D]{Tuck School of Business,
				Dartmouth College,
				\printead{e5}}
		\end{aug}
		
		\begin{abstract}
			It is widely speculated that auditors' public forecasts of bankruptcy are, at least in part, self-fulfilling prophecies in the sense that they actually cause bankruptcies that would not have otherwise occurred. This conjecture is hard to prove, however, because the strong association between bankruptcies and bankruptcy forecasts could simply indicate that auditors are skillful forecasters with unique access to highly predictive covariates. 
			In this paper, we investigate the causal effect of bankruptcy forecasts on bankruptcy using nonparametric sensitivity analysis. We contrast our analysis with two alternative approaches:  a linear bivariate probit model with an endogenous regressor, and a recently developed bound on risk ratios called E-values. Additionally, our machine learning approach incorporates a monotonicity constraint corresponding to the assumption that bankruptcy forecasts do not make bankruptcies less likely. Finally, a tree-based posterior summary of the treatment effect estimates allows us to explore which observable firm characteristics moderate the inducement effect.
		\end{abstract}
		
		
	\end{frontmatter}
	\section{Introduction}
	A ``going concern opinion'' is an assessment by an auditor that a firm is at risk of going out of business in the coming year. Here, a ``concern'' refers to a firm, and ``going'' refers to staying, as opposed to going out of, business. According to U.S.~securities regulations, a public company that receives an adverse going concern opinion must disclose it in the firm's annual filings with the Securities and Exchange Commission. Once issued and disclosed, a going concern opinion may directly contribute to a firm's bankruptcy risk, for example, by inducing lenders to pull lines of credit or increase borrowing costs.\footnote{See \cite{maurer-wsj-2020} for a recent discussion of going concern opinions in the news. See \cite{Chen-He-Ma-etal-2016} for a discussion of how adverse going concern opinions can adversely affect borrowing costs.} As reported in \cite{maurer-wsj-2020}:
	
	\begin{quote}
		Companies that receive a going-concern audit opinion may be subjected to more rigorous covenant terms or downgrades in their credit ratings, said Anna Pinedo, a partner at law firm Mayer Brown. Fractured relationships with customers could also strengthen a business’s competitors, she said.
	\end{quote}
	
	Estimating the magnitude of such an ``inducement effect'' is complicated by the unavailability of the auditors' private information to the analyst. That is, in addition to publicly available firm information, auditors have access to ``private information'' gleaned from confidential documents and via firsthand knowledge of undocumented attributes such as the firm's corporate culture. This paper considers the question: do going-concern opinions help to predict bankruptcy because they incorporate the auditor's private information or because of an inducement effect? This is a textbook example of causal inference where the potential unobserved confounders are particularly pictureseque: what do auditors know that we (the analysts) do not? We introduce methodology to quantify the impact of private information on the probability that a firm files for bankruptcy in the fiscal year following the issuance of a going concern opinion.  We conduct a sensitivity analysis rooted in nonlinear, semiparametric regression techniques and a generalization of the bivariate probit model with an endogenous regressor. Our use of ``machine learning'' tools to study this problem adds to a growing literature on applying machine learning methods to accounting data.  \cite{bao2020detecting} deploy an ensemble model to predict fraud, \cite{brown2020you} incorporate a Bayesian topic modeling algorithm to predict intentional financial misreporting, and  \cite{bertomeu2021using} provide an overview of how machine learning methods are growing in accounting research, specifically in regards to the study of accounting misstatements. We conclude that there is evidence for inducement under plausible assumptions on the distribution of the auditors' private information.

	\subsection{Methodological background}
	
	Denote the treatment variable by $G_i$ for ``going concern'' so that $G_i = 1$ for the $i$th firm in our sample if that firm received an adverse going concern opinion in the prior year. Denote the outcome variable $B_i$ for ``bankrupt'' so that $B_i = 1$ filed for bankruptcy.
	In terms of potential outcomes \citep{rubin}, we are interested in two scenarios: $B^1_i$ and $B^0_i$, which are the outcome of a firm $i$ if it had received the treatment and if it had not received the treatment; only one of these potential outcomes is observed. 
	
	The primary estimand of interest will be the causal risk ratio (CRR):
	\begin{equation}
		\tau \equiv \E(B^1)/\E(B^0)
	\end{equation}
	which we will often refer to as simply the ``inducement effect.'' Alternatively, we can define the inducement effect in terms of the ``do''-operator of \cite{pearl-2000} as 
	\begin{equation}
		\tau \equiv \E(B=1 \mid \text{do}(G=1))/\E(B=1 \mid \text{do}(G=0))
	\end{equation}
	where $\text{do}(G=g)$ refers to an exogeneous intervention, in contradistinction to probabilistic conditioning. We will also consider the risk difference
	\begin{equation}
		\Delta \equiv \E(B^1) - \E(B^0)
	\end{equation}
	and consider how these two estimands differ as a function of observable firm characteristics.
	
	The fundamental problem of causal inference \citep{Holland-1986} is that $(B^1, B^0)$ are never observed simultaneously, rather only one or the other is observed. Consequently, the conditions under which the CRR can be estimated must be carefully assessed and their plausibility debated. There are three widely used methods for estimating average treatment effects: randomization, regression adjustment (broadly construed to include matching and propensity score based methods), and instrumental variables analysis. To briefly review:
	\begin{itemize}
		\item In a randomized controlled trial, the treatment variable---$G$ in the present context---is independent of the potential outcomes $B^1$ and $B^0$; in this case, $\E(B^1) = \E(B \mid G = 1)$, the right hand side of which is readily estimable from observed data (and likewise for the $G = 0$ case). 
		
	\item When a randomized experiment is not possible (such as in the present example) one instead may hope to find a set of control variables $\mathbf{x}$ for which $\E(B^1 \mid \mathbf{x}) = \E(B \mid G = 1, \mathbf{x})$ and $\E(B^0 \mid \mathbf{x}) = \E(B \mid G = 0, \mathbf{x})$, in which case treatment effects can be estimated by estimating these conditional expectations via regression modeling. This condition is called {\em conditional ignorability}, or, alternatively, $\mathbf{x}$ are said to satisfy the {\em back-door criterion} \cite{pearl-2000}. \color{black}
		
		\item A third possibility is that a sufficient set of controls is unavailable, but an {\em instrument} for the treatment assignment is available. An instrument is a variable that is causally related to the treatment but not otherwise associated with the response variable. In the current context, an instrument variable (IV) would be a one that affects the probability that an auditor issues a going concern opinion without directly affecting bankruptcy probabilities or sharing common causes with bankruptcies. 
		Here, we do not elaborate on the details of instrumental variable regression, but see \cite{imbens2014instrumental} for a recent survey and \cite{larcker2010use} for a discussion of the use of IV specifically in accounting research.
	\end{itemize}
	In the present context, none of these three approaches are available. A sufficient set of controls is certainly not readily available and the existence of a valid instrument is doubtful because firms choose their own auditor, rendering auditor attributes endogenous. Although there are other approaches---such as regression continuity design \citep{Imbens-rdd,Thistlethwaite-rdd}, difference-in-differences \citep{card1994}, and the synthetic control method \citep{abadie2010,abadie2003}---they apply in idiosyncratic settings that are not representative of the bankruptcy inducement problem. 
	
	With none of the usual tools available to us, it may be possible to make additional modeling assumptions that yield identification of the treatment effect. One such model for bivariate binary observations is the bivariate probit model with an endogeneous regressor \citep[Section~ 15.7.3]{Wooldridge-2010}. Such model-based identification is generally undesirable because the identifying form of the likelihood typically lacks plausible justification \citep{Manski}. Accordingly, it is prudent to consider a range of different assumptions (model specifications) and observe how the estimated treatment effects vary as a result. In this paper, we propose a method for modeling the strength of unobserved confounding in a machine learning framework which permits convenient sensitivity analysis without unrealistically constraining the observed data distribution.
	
	\subsection{Methodological contribution of this paper}
	
	This paper brings together three lines of methodological research. First, we develop a generalization of the bivariate probit with endogeneous regressor and use this unidentified model to conduct a sensitivity analysis. Second, we use modern Bayesian tree-based classification models to estimate the identified parameters in our model and describe a numerical procedure to map these parameters back to the causal estimands of interest. This approach represents both a novel use of Bayesian machine learning as well as a novel application of machine learning to the applied problem of whether going concern opinions induce bankruptcy. Additionally, this model incorporates the assumption that going concern opinions cannot make bankruptcies less likely, a plausible assumption that potentially improves estimation accuracy. Finally, we apply a tree-based posterior summarization strategy to our estimates of the individual treatment effects to identify interesting subgroups for further scrutiny, a method first described in \cite{bcf}, building on a framework laid out in \cite{Hahn-2015} for linear models.
	
	\subsection{Paper structure}
	
	Because this work touches on many disparate areas, an overview organizing the contents may be helpful.
	\begin{itemize}
		\item First, we review the traditional parametric model used for the binary-treatment-binary-response setting with unmeasured confounding, which is the bivariate probit model with endogenous regressor. We provide a novel justification of this model in terms of Pearl's causal calculus using a latent factor representation of the bivariate probit likelihood. 
		\item Next, we generalize this model by relaxing the linearity and distributional assumptions, making it robust to misspecification.
		\item The generalized bivariate probit model is not point identified, making a sensitivity analysis necessary. A computationally efficient method for conducting the sensitivity analysis is developed, which uses a single Bayesian model fit of the reduced form parameters.
		\item We then introduce monotone Bayesian additive regression trees, which is a custom modification of the popular BART model \citep{bart}, and describe the Markov chain updates for enforcing monotonicity in the treatment variable.
		\item Putting these pieces together, the new machine learning sensitivity analysis is applied to over 20,000 data points from publicly traded U.S. firms. Results are compared to a model-free sensitivity analysis approach called E-values \citep{Peng-2016}, which generalize the well known Cornfield bounds \citep{Cornfield}. Decision trees are used as a posterior summarization tool to discover variables that moderate the inducement effect.  
		\item Additionally, the new approach is investigated via several simulation studies to evaluate its behavior relative to alternative approaches when the data generating process is known.
	\end{itemize}
	
	\section{The bivariate probit model with endogenous predictor}
	
	A well-known model that has been used for problems similar to the one described here is the bivariate probit with endogenous predictor \citep[Section~ 15.7.3]{Wooldridge-2010}. This model can be expressed in terms of bivariate Gaussian latent utilities $Z_g$ and $Z_b$ that relate to going concern opinions and bankruptcy: 
	\begin{equation}
		\begin{pmatrix}
			Z_{g,i}\\
			Z_{b,i}
		\end{pmatrix}\stackrel{\text{iid}}{\sim}\mathcal{N}(\boldsymbol{\mu}, \bm{\Sigma})
		\qquad \boldsymbol{\mu}=\begin{pmatrix}
			\beta_0+\beta_1\bm{x}_i\\
			\alpha_0+\alpha_1\bm{x}_i
		\end{pmatrix}
		\qquad
		\bm{\Sigma}=\begin{pmatrix}
			1&\rho\\
			\rho&1
		\end{pmatrix}.
		\label{latentutility}
	\end{equation}
	The premise of this model is that $\rho$ reflects the influence of private information available to the auditor but not the researcher, and $\bm{x}_i$ represents covariates of a company that is available to both the auditor and to the researcher. The observed binary indicators, $G$ and $B$, relate to these latent utilities via
	\begin{align}
		G&=\mathbbm{1}\cbr{Z_{g,i}\geq 0}\\
		B&=\mathbbm{1}\cbr{Z_{b,i}\geq -\gamma G}
		\label{align1}
	\end{align}
	The coefficient $\gamma$ governs the strength of the inducement effect.
	
\begin{figure}[h]
		\centering
		\begin{subfigure}{.4\textwidth}
			\includegraphics[scale=0.44]{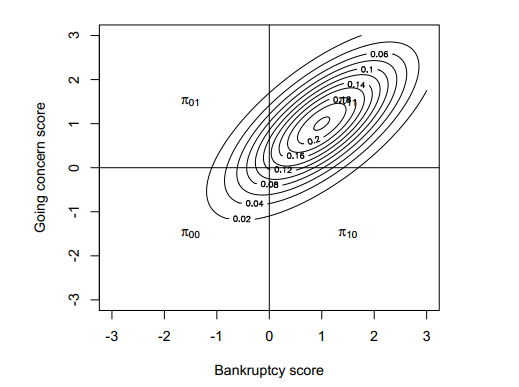}
		\end{subfigure}
		\begin{subfigure}{.4\textwidth}
			\includegraphics[scale=0.515]{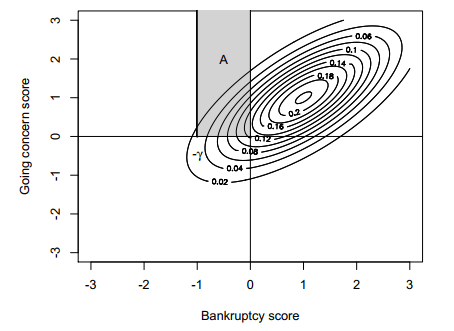}
		\end{subfigure}
		\caption{The bivariate probit entails ellipse shaped probability contours, where (when $\gamma=0$) the probability mass associated to each quadrant represents the four combinations of the bivariate binary observed variables $(B, G)$. The shaded region in the right panel, labeled ``A'', is subtracted from the upper left quadrant and added to the upper right quadrant when a going concern is issued, thus reflecting the endogeneity of the going concern variable. The parameters $\rho$ and $\gamma$ are estimable because changes in the shape of the ellipses, governed by $\rho$, lead to more distinct apportioning of probability than do changes in the width of the A region, governed by $\gamma$.}
		\label{ellipse}
	\end{figure}

	The basic identification strategy can be motivated geometrically. Let $$\bm{\Pi}=\begin{pmatrix}
		\pi_{01}&\pi_{11}\\
		\pi_{00}&\pi_{10}
	\end{pmatrix}$$
	where $\pi_{jk}=\Pr(B=j, G=k)$, which describes the four scenarios resulting from our equations for $G$ and $B$.  \autoref{ellipse} gives a visual representation of the $\bm{\Pi}$ matrix.
	
	Note in \autoref{ellipse} that $\boldsymbol{\mu}$ determines the location (center of ellipse) and the correlation $\rho$ determines the tilt and concentration of the probability contours. Inducement introduces an extra parameter which lowers the threshold for bankruptcy by $\gamma$.  
	
	\subsection{A causal interpretation of $\gamma$}
	
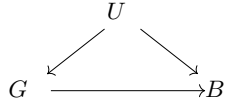
\begin{figure}[h!]
		\begin{tikzcd}
			& U \arrow[dr] \arrow[dl]\\
			G \arrow[rr] && B
		\end{tikzcd}
		\caption{Conditional on observable attributes $\mathbf{x}$ (not shown), the causal diagram above stipulates the temporal ordering among the firm's private information $U$, the auditor risk assessment $G$, and the firm's bankruptcy outcome, $B$.}\label{causaldiag} 
	\end{figure}

	Having presumed a particular parametric model for the distribution of the data $(G,B)$ (conditional on covariates $\mathbf{x}$), we would like additional license for the interpretation that $\rho$ captures the contribution of auditor's additional information on bankruptcy likelihood while $\gamma$ captures the contribution of inducement effects on bankruptcy likelihood. To justify this interpretation, we turn to the causal analysis framework of \cite{pearl-2000}. Recall that in Pearl's framework, the inducement effect would be written as 
	\begin{equation}\label{docalc}
		\mbox{Pr}(B = 1 \mid \mathbf{x}, \text{do}(G = 1)) / \mbox{Pr}(B = 1 \mid \mathbf{x}, \text{do}(G = 0)),
	\end{equation}
	where $\text{do}(G=1)$ denotes the intervention of issuing a going concern, irrespective of the stochastic data generating process. Denote by $U$ the auditor's additional information.  Suppressing the covariates $\mathbf{x}$, the relationship between $G$ and $B$ can be expressed using the causal diagram depicted in \autoref{causaldiag}.
	
	This diagram asserts several causal assumptions.  First, the issuance of a going concern does not cause the existence of auditor's additional information:  there is no arrow running from $G$ to $U$.  Second, bankruptcies cannot cause going concerns:  there is no arrow running from $B$ to $G$.  Similarly, bankruptcies do not cause the creation of auditor's additional information for predicting bankruptcy:  there is no arrow from $B$ to $U$.  All of these assumptions follow straightforwardly from a temporal ordering---auditors first procure information concerning bankruptcy propensity ($U$), they then issue going concern opinions ($G$), and then firms either go bankrupt or not ($B$).  
	
	Because $U$ disconnects alternative routes from $B$ to $G$ and no directed path exists from $G$ to $U$, $U$ is said to satisfy the back-door criterion \cite{pearl-2000}, and we can compute $\mbox{Pr}(B = 1 \mid \mathbf{x}, \text{do}(G = 1))$ via the expression:
	\begin{equation}\label{backdoor}
		\mbox{Pr}(B = 1 \mid \mathbf{x}, \text{do}(G = 1)) =  \int  \mbox{Pr}(B = 1 \mid \mathbf{x}, U=u, G = 1) f(u) \dd u,
	\end{equation}
	where $f(u)$ is the marginal density of the random variable $U$.
	
	The difficulty, of course, is that $U$ is unobserved in our problem so $f(u)$ can never be estimated from data.  However, we can re-express the bivariate probit model directly in terms of $U$ in order to derive the  expression  of \autoref{backdoor} in terms of parameters $\rho$, $\gamma$, and $\beta$.  This demonstrates how the functional form of the model dictates the causal estimand in \autoref{docalc}, which in turn establishes the causal interpretation of the $\gamma$ parameter.  
	
	In detail, re-writing \autoref{latentutility} conditional on $U$ gives a model with diagonal error covariance:
	\begin{equation}
		\begin{split}
			\begin{pmatrix}
				Z_{g,i} \\ Z_{b,i}
			\end{pmatrix} &\sim \mbox{N}(\boldsymbol{\mu}, \bm{\Sigma}), \;\;\;\; 
			\boldsymbol{\mu}  = \begin{pmatrix}
				\beta_{0} + \beta_{1}\mathbf{x}_i + \eta_g U  \\
				\alpha_{0} + \alpha_{1}\mathbf{x}_{i}  + \eta_b U
			\end{pmatrix},  \;\;\;\;  \bm{\Sigma} = \begin{pmatrix} v_g & 0 \\ 0 & v_b \end{pmatrix},
		\end{split}
	\end{equation}
	where $U \sim \mbox{N}(0, 1)$, $v_g = 1 - \eta_g^2$, $v_b = 1 - \eta_b^2$ and $\rho = \eta_g \eta_b$.  Although this representation is non-unique in $(\eta_g, \eta_b, v_g, v_b)$, it turns out that the expression in \autoref{backdoor} will not depend on these values.  This representation allows us to apply the causal assumptions depicted in the causal diagram above, which in turn allows us to derive the counterfactual probability of bankruptcy as:
	\begin{equation}
		\begin{split}
			\mbox{Pr}(B = 1 \mid \mathbf{x}, \text{do}(G = 1)) &=  \int  \mbox{Pr}(B = 1 \mid \mathbf{x}, U=u, G = 1) \N_u(0,1) \dd u,\\
			& = \int  1 - \Phi(0; \gamma + \alpha_0 + \alpha_1 \mathbf{x} + \eta_b u) \N_u(0,1) \dd u,\\
			& = \int 1 - \int_{-\infty}^0 \N_w(\gamma + \alpha_0 + \alpha_1 \mathbf{x} + \eta_b u, v_b) \dd w \N_u(0,1) \dd u,\\
			& = 1 - \int_{-\infty}^0 \int \N_w(\gamma + \alpha_0 + \alpha_1 \mathbf{x} + \eta_b a, v_b)  N_u(0,1) \dd a \dd w,\\
			& = 1 - \int_{-\infty}^0 \N_w(\gamma + \alpha_0 + \alpha_1 \mathbf{x}, 1) \dd w,\\
			& = 1 - \Phi(0; \gamma + \alpha_0 + \alpha_1 \mathbf{x}),\\ 
			&= \Phi(\gamma + \alpha_0 + \alpha_1 \mathbf{x}).
		\end{split}
	\end{equation}
	Here $\Phi(0; \mu)$ denotes the CDF of a normal distribution with mean $\mu$ and variance 1, evaluated at 0.  A similar calculation can be done for $\mbox{Pr}(B = 1 \mid \mathbf{x}, \text{do}(G = 0))$, allowing us to recover the causal risk ratio as $$\tau(\mathbf{x}_i) = \Phi(\gamma + \alpha_0 + \mathbf{x}\alpha_1) /\Phi(\alpha_0 + \mathbf{x}\alpha_1).$$  In other words, fitting a bivariate probit model to the data $(G, B, \mathbf{x})$, coupled with the causal assumptions encoded in the causal diagram \autoref{causaldiag}, implies a causal inducement effect that can be written in terms of $\alpha$ and $\gamma$. Although $\gamma$ is a shared constant parameter, its impact on the risk ratio for a given firm will depend on both $\mathbf{x}$ and $\alpha$. 
	
	\subsection{Identification and estimation for bivariate probit models}\label{techremarks}
	
	The previous section related the parameters of the bivariate probit model with endogenous regressor to the causal risk ratio. However, identifiability is a distinct concern. Identification of parameters in bivariate probit models is subtle and deserves a careful discussion. The treatment in \cite{Heckman-1978} derives the bivariate probit model from a system of simultaneous equations.  Section 3 of \cite{Heckman-1978}, page 949, provides a proof that the associated reduced form parameters of the model are identified without any exclusion restrictions, which would require that the going concern and bankruptcy equations do not share all of their covariates in common.  Identification follows from the functional form of the probit likelihood, and indeed \cite{Heckman-1978} contains a section devoted to maximum likelihood estimation.  \cite{Heckman-1978} also treats the continuous (non-binary response) version of the same structural system; in that case, exclusion restrictions are necessary for identification, and, in that case, estimation can proceed by a two-stage least squares procedure without specifying a likelihood function.  
	
	\cite{Evans-Schwab-1995} study an applied problem using the binary response formulation of the \cite{Heckman-1978} model, but do not assume the probit formulation and rather proceed to estimate parameters using an OLS based procedure.  In this context, the role of an exclusion restriction is an open question as  \cite{Altonji-Elder-Taber-2005} point out; however, the two-step procedure applied to the binary response setting gives inconsistent estimates. 
	
	In summary, textbook treatments of the bivariate probit model equivocate on the necessity of an exclusion restriction \citep[Chapter~15]{Wooldridge-2010}.  To be clear, if one assumes the bivariate probit formulation, then an exclusion restriction is not necessary.  If fitting a generalized linear model to a bivariate binary response {\em without} specifying a link function, an exclusion restriction is necessary. Here, these concerns are secondary, as we do not demand identification, but proceed instead via a sensitivity analysis.
	
	\section{Modular sensitivity analysis with machine learning}\label{section_indirect_inference}
	
	In this section we propose our new approach for machine learning-based sensitivity analysis by generalizing the bivariate probit model.  We begin by defining the joint probability of treatment and outcome as
	\begin{equation}
		\Pr\left(B, G\mid \mathbf{x}\right)=\int_{\mathbb{R}} \Pr\left(B\mid \mathbf{x},U=u,G\right)\Pr\left(G\mid \mathbf{x},U=u\right)f(u)\dd u
		\label{eq2}
	\end{equation}
	for latent variable $U$. In this formulation, $U$ has two special properties. First, it is assumed to be the {\em orthogonal} component of the private information in the sense that $U \independent X$, hence $\mathbf{x}$ does not appear in $f(u)$. Second, $U$ is assumed to be {\em complete}, in the sense that $\Pr(B \mid \mathbf{x}, u, G)$ can be interpreted causally in $G$, because $U$ is a sufficient control variable. That is, $\Pr(B^1 \mid \mathbf{x}, u) = \Pr(B \mid \mathbf{x}, \text{do}(G = 1), u) = \Pr(B \mid \mathbf{x}, G = 1, u)$ and similarly for $G = 0$; accordingly, the inducement effect for firm $i$ is
	\begin{equation}
		\tau(\mathbf{x}_i)\equiv\frac{\int_{\mathbb{R}}\Pr\left(B=1\mid \mathbf{x},G=1,u\right)f(u)\dd u }{\int_{\mathbb{R}}\Pr\left(B=1\mid \mathbf{x},G=0,u\right)f(u)\dd u}.
		\label{realtreat}
	\end{equation}
	Because the outcome and treatment are both binary, we can expand this probability into its four constituent parts.  For convenience, we specify a probit link, yielding 
	\begin{align}
		\begin{split}
			\Pr\left(B=1\mid \mathbf{x},U=u,G=1\right)& = \Phi\left(b_1(\mathbf{x})+u\right),\\
			\Pr\left(B=1\mid \mathbf{x},U=u,G=0\right)&= \Phi\left(b_0(\mathbf{x}\right)+u), \\
			\Pr\left(G=1\mid \mathbf{x},U=u\right)&= \Phi\left(g(\mathbf{x})+u\right).
		\end{split}
		\label{maineq}
	\end{align}
	Therefore, in terms of $f$, $b_1$, $b_0$ and $g$, the individual inducement effect for firm $i$ is
	\begin{align}
		\tau(\mathbf{x}_i)=\frac{\int_{\mathbb{R}} \Phi\qty(b_1(\mathbf{x})+u)f(u)\dd u }{ \int_{\mathbb{R}} \Phi\qty(b_0(\mathbf{x})+u) f(u)\dd u}
		\label{treateq}
	\end{align}
	and we denote the sample average inducement effect (or average causal risk ratio: ACRR) as $\bar{\tau} = \frac{1}{n}\sum_{i=1}^{n}\tau(\mathbf{x}_i)$. Importantly, the orthogonality and completeness of $U$, as well as the choice of the probit link, are not substantive assumptions, as $U$ is unobserved and $b_1$, $b_0$ and $g$ are nonparametric functions of $\mathbf{x}$. Rather, these assumptions {\em define} $U$ and give the specification of $f(\cdot)$ meaning; the choice of $f$, therefore, {\em is} a substantive assumption (as it is in the bivariate probit model as well).
	
	This formulation entails that as $u\rightarrow -\infty$, the probability of bankruptcy approaches 0, regardless of whether the treatment is administered or not. As $u\rightarrow \infty$, the probability of bankruptcy approaches 1.  The special case $u=0$ corresponds to no unobserved confounding and the inducement effect can be computed directly from the observed joint probabilities.      	
	Finally, because $G$ and $B$ must have a valid joint distribution at each $\mathbf{x}$ value, we have the following system of equations defining our data generating process:
	\begin{equation}
		\begin{split}
			\Pr\qty(B=1, G=1\mid \bm{x})&= \int_{\mathbb{R}} \Phi\qty(g(\bm{x})+u)\Phi\qty(b_1(\bm{x})+u)f(u)\dd{u},\\
			\Pr\qty(B=1, G=0\mid \bm{x})&=\int_{\mathbb{R}} \qty(1-\Phi\qty(g(\bm{x})+u))\Phi\qty(b_0(\bm{x})+u)f(u)\dd{u},\\
			\Pr\qty(B=0, G=1\mid \bm{x})&=\int_{\mathbb{R}}\Phi\qty(g(\bm{x})+u)\qty(1-\Phi\qty(b_1(\bm{x})+u))f(u)\dd u.
			\label{long}
		\end{split}
	\end{equation}
	Observe that this generalizes the bivariate probit model with endogenous regressor: when $U \sim \mbox{N}(0, \rho/(1-\rho))$,  $b_0(\mathbf{x}) = \alpha_0 + \alpha_1 \mathbf{x}$, $b_1(\mathbf{x}) = \alpha_0 + \alpha_1 \mathbf{x} + \gamma$, and $g(\mathbf{x}) = \beta_0 + \beta_1 \mathbf{x}$ we recover that model exactly.	Our formulation is quite a lot more flexible: we relax the Gaussian assumption on the marginal distribution of $U$, drop the parallel relationship between $b_0(\cdot)$ and $b_1(\cdot)$, and allow $b_1$, $b_0$ and $g$ to be nonlinear.\footnote{Observe that when the form of $b_1$, $b_0$ and $g$ are constrained, as in the linear probit model, the choice of the probit link becomes a substantive modeling assumption, while in our more flexible formulation it is merely a convenience.} The price of the extra flexibility of our relaxed specification is that $f(u)$ is now unidentified, whereas in the bivariate probit case it is assumed to be Gaussian but with an identified correlation parameter $\rho$. 
	
	The left hand side of the system in \autoref{long}---the {\em reduced form} parameters---can be estimated from the observed data.  Any of a host of machine learning classification methods, such as random forest \citep{rf}, xgboost \citep{boost}, Bayesian additive regression trees (BART) \citep{bart}, among others, can be used to obtain estimates of these probabilities. Here, we focus our attention on BART for two reasons: one, we can impose monotonicity so that going concerns can only increase the probability of bankruptcy, and two, we obtain a Bayesian measure of uncertainty based on Markov chain Monte Carlo sampling methods. 
	
	\subsection{Projecting the reduced form probabilities onto the causal parameters}
	
	What remains is to solve for $b_1(\cdot),b_0(\cdot), g(\cdot)$, the {\em structural}, or causal, parameters. To do so, we take a numerical approach, by minimizing  the sum of the squared distance between the three left-hand right-hand pairs in \autoref{long}:
	\begin{align*}
		&\left[\Phi^{-1}\left(\Pr\qty(B=1, G=1\mid \mathbf{x})\right)-\Phi^{-1}\left(\int_{\mathbb{R}} \Phi\qty(g(\mathbf{x})+u)\Phi\qty(b_1(\mathbf{x})+u)f(u)\dd{u}\right)\right]^2+\\
		&\left[\Phi^{-1}\left(\Pr\qty(B=1, G=0\mid \mathbf{x})\right)-\Phi^{-1}\left(\int_{\mathbb{R}} \qty(1-\Phi\qty(g(\bm{x})+u))\Phi\qty(b_0(\bm{x}+u))f(u)\dd{u}\right)\right]^2+\\
		&\left[\Phi^{-1}\left(\Pr\qty(B=0, G=1\mid \mathbf{x})\right)-\Phi^{-1}\left(\int_{\mathbb{R}}\Phi\qty(g(\bm{x})+u)\qty(1-\Phi\qty(b_1(\bm{x})+u))f(u)\dd u\right)\right]^2.
	\end{align*}
	
	Although it is unclear that \autoref{long} has a unique solution in $b_1$, $b_0$, $g$, numerical solvers converge readily in our experience. Heuristically, as a convex combination of monotone functions, each of the individual integrals in \autoref{long} is likely to be nearly linear over much of its domain. Note that the use of the normal inverse CDF simply ensures that the range of our objective function is unbounded; we observe that this improves numerical stability of our solver.
	
	We refer to this process as {\em modular} because it requires fitting the reduced form model just one time. Sensitivity of the causal estimates to different choices of $f$ can be assessed independently using the same estimates (or posterior samples) from a single reduced form model fit.
	
	\section{Monotone BART for reduced form inference}
	
	\subsection{Probit BART Overview}
	
\begin{figure}
		\begin{center}
			\begin{tikzpicture}[
				scale=0.8,
				node/.style={%
					draw,
					rectangle,
				},
				node2/.style={%
					draw,
					circle,
				},
				]
				\node [node] (A) {$x_1<0.8$};
				\path (A) ++(-135:\nodeDist) node [node2] (B) {$m_{l1}$};
				\path (A) ++(-45:\nodeDist) node [node] (C) {$x_2<0.4$};
				\path (C) ++(-135:\nodeDist) node [node2] (D) {$m_{l2}$};
				\path (C) ++(-45:\nodeDist) node [node2] (E) {$m_{l3}$};
				
				\draw (A) -- (B) node [left,pos=0.25] {no}(A);
				\draw (A) -- (C) node [right,pos=0.25] {yes}(A);
				\draw (C) -- (D) node [left,pos=0.25] {no}(A);
				\draw (C) -- (E) node [right,pos=0.25] {yes}(A);
			\end{tikzpicture}
			\hspace{0.1\linewidth}
			\begin{tikzpicture}[scale=3]
				\draw [thick, -] (0,1) -- (0,0) -- (1,0) -- (1,1)--(0,1);
				\draw [thin, -] (0.8, 1) -- (0.8, 0);
				\draw [thin, -] (0.0, 0.4) -- (0.8, 0.4);
				\node at (-0.1,0.4) {0.4};
				\node at (0.8,-0.1) {0.8};
				\node at (0.5,-0.2) {$x_1$};
				\node at (-0.3,0.5) {$x_2$};
				\node at (0.9,0.5) {$m_{l1}$};
				\node at (0.4,0.7) {$m_{l2}$};
				\node at (0.4,0.2) {$m_{l3}$};
			\end{tikzpicture}
		\end{center}
		\caption{(Left) An example binary tree, with internal nodes labelled by their splitting rules and terminal nodes labelled with the corresponding parameters $m_{lb}$. (Right) The corresponding partition of the sample space and the step function.  Figure from \cite{bcf}.}
		\label{fig:treestep}
	\end{figure}
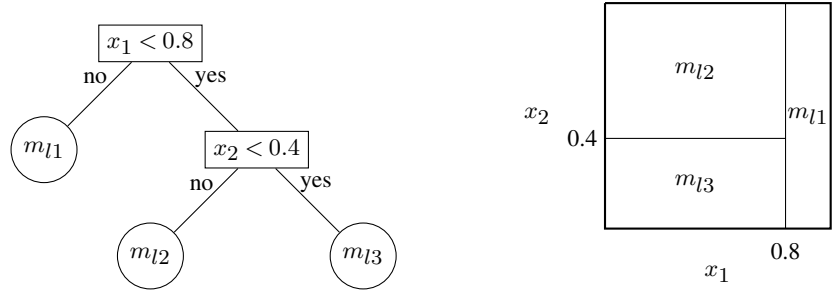

	BART, Bayesian additive regression trees, is at its core a sum-of-trees model. For a $p$-dimensional vector of covariates $\bm{x}$ and a continuous response variable $Y$, the BART model is 
	\begin{equation}
		Y=t(\bm{x})+\varepsilon, \qquad \varepsilon\sim \N(0, \sigma^2)
		\label{bart_setup}
	\end{equation}
	where $t(\bm{x})=\E(Y\mid \bm{x})$ denotes a sum of $L$ regression trees (i.e., $t(\bm{x})=\sum_{l=1}^{L} q_l(\bm{x})$). \autoref{fig:treestep} presents an example regression tree.  In addition to this additive tree representation, BART uses a stochastic process tree prior that favors smaller trees; the prior probability of splitting at depth $d$ is $\eta(1+d)^{-\zeta}, \; \eta\in (0,1), \; \zeta\in[0, \infty)$ \citep{Chipman-1998}.    
	
	At each leaf of the tree, parameters are assigned independent regularization priors, $m_{lb}\sim  \N(0, \sigma_{\mu}^2)$, where $\sigma_{\mu}=0.5/(k\sqrt{L})$, and $L$ is the number of trees. 
	
	To handle binary outcomes, BART may be extended through a latent probit formulation, using the data augmentation approach of \cite{Albert-1993}. For binary outcome $B$: 
	\begin{align*}
		&B^*=t(\bm{x})+\varepsilon, \qquad \varepsilon\sim \N(0,1),\\
		&B=\mathbbm{1}(B^*>0),
	\end{align*}
	which implies
	\begin{equation}
		\Pr(B=1\mid \bm{x})=\Phi(t(\bm{x}))
		\label{bart_binary}
	\end{equation}
	where $\Phi$ is the standard normal CDF.  
	
	The $B^*$ variables may be imputed from their truncated normal full conditional distributions; conditional on $B^*$ the BART fitting algorithm can be applied as usual.
	
	\subsection{Monotone probit BART}
	
	We turn now to a modification of the BART probit model for the bankruptcy and going concern data. We model the left-hand side of the system in \autoref{long} using a compositional representation, using two ``chained'' regression models, one for $\Pr(G \mid \mathbf{x})$ and another for $\Pr(B \mid \mathbf{x}, G)$. This formulation permits us to insist that $\Pr(B=1\mid G=1, \mathbf{x}) \geq \Pr(B=1\mid G=0, \mathbf{x})$ for all $\mathbf{x}$, encoding the uncontroversial belief that adverse going concern opinions never mitigate bankruptcy risk. To enforce this constraint, we parameterize $\Pr(B=1\mid G, \mathbf{x})$ as follows:
	\begin{equation}
		\begin{split}
			\Pr(B=1\mid G=1, \mathbf{x}) &= \Phi[h_1(\mathbf{x})],\\
			\Pr(B=1\mid G=0, \mathbf{x}) &= \Phi[h_0(\mathbf{x})]\Pr(B=1\mid G=1, \mathbf{x}),
			\\ &= \Phi[h_0(\mathbf{x})]\Phi[h_1(\mathbf{x})],\\
			\Pr(G=1\mid \mathbf{x}) &= \Phi[w(\mathbf{x})].
		\end{split}
	\end{equation}
	For each function $h_0, h_1,$ and $w$ we specify independent BART priors which allows us to fit the treatment and outcome models separately. 
	
	The likelihood for the bankruptcy model is
	\begin{equation}
		\begin{split}
			L(h_0, h_1; B, G, \mathbf{X}) =&\prod_{i: G_i=1}\Phi(h_1(\mathbf{x}_i))^{B_i}(1-\Phi(h_1(\mathbf{x}_i)))^{1-B_i} \times \\
			&\prod_{i: G_i=0}[\Phi(h_0(\mathbf{x}_i))\Phi(h_1(\mathbf{x}_i))]^{B_i} (1-\Phi(h_0(\mathbf{x}_i))\Phi(h_1(\mathbf{x}_i)))^{1-B_i}\label{eq:bart-da0}.
		\end{split}
	\end{equation}
	This likelihood is challenging: The expression $1- \Phi(h_0(\mathbf{x}_i))\Phi(h_1(\mathbf{x}_i))$ does not factor into separate terms involving the unknown functions $h_0$ and $h_1$, making it difficult to adapt the BART MCMC sampler for posterior inference. To overcome this challenge, we introduce a data-augmented representation that permits updating $h_0$ and $h_1$ independently using standard MCMC for probit BART. 
	
	To begin, note that the first term above (corresponding to $G=1$) involves only $h_1$ so we only need to augment data in the $G=0$ ``arm.'' When $G=0$, we relate $B$ to two independent binary latent variables $R_0$ and $R_1$  as follows:
	\begin{gather*}
		\Pr(R_0=1\mid\mathbf{x}, G=0) = \Phi(h_0(\mathbf{x})),\\
		\Pr(R_1=1\mid\mathbf{x}, G=0) = \Phi(h_1(\mathbf{x}))\label{eq:Rprobs}
	\end{gather*}
	and $B = R_0R_1$. Integrating out the latent variables gives $\Pr(B=1\mid\mathbf{x}, G=0) = \Phi(h_0(\mathbf{x}))\Phi(h_1(\mathbf{x}))$ and $\Pr(B=0\mid\mathbf{x}, G=0) = 1-\Phi(h_0(\mathbf{x}))\Phi(h_1(\mathbf{x}))$ as required.\footnote{Observe that $\Pr(B=1\mid\mathbf{x}, G=1) = \Pr(R_1=1\mid\mathbf{x}, G=0)$, so thinking about this as a generative model we can interpret $R_1$ as a simulated outcome if we had observed $G=1$ and $R_0$ as an indicator that this outcome is ``thinned'' to enforce monotonicity, because, in reality, $G=0$.} 
	The augmented likelihood function (including $R_0, R_1$) is 
	\begin{equation}
		\begin{split}
			L(h_0, h_1; R, B, G, \mathbf{X}) =&\prod_{i: G_i=1}\Phi(h_1(\mathbf{x}_i))^{B_i}(1-\Phi(h_1(\mathbf{x}_i)))^{1-B_i} \times \\
			&\prod_{i: G_i=0} \Phi(h_1(\mathbf{x}_i))^{R_{1i}}(1-\Phi(h_1(\mathbf{x}_i)))^{1-R_{1i}}\times \\
			&\prod_{i: G_i=0}\Phi(h_0(\mathbf{x}_i))^{R_{0i}}(1-\Phi(h_0(\mathbf{x}_i)))^{1-R_{0i}}\times\\
			&\prod_{i: G_i=0}\ind{B_i = 1\text{ if }R_{0i}=R_{1i}=1} 
			\label{eq:bart-da1}
		\end{split}
	\end{equation}
	After rearranging terms, we have two separate probit likelihoods in $h_0$ and $h_1$ (and the domain restriction in the last term). Conditional on $R_0, R_1$ we can update $h_0, h_1$ using standard probit BART MCMC steps. To update the latent variables $R_{0i}$ and $R_{1i}$, first note that they are fixed at 1 when $B_i=1$ and $G_i=0$. When $B_i=0$ and $G_i=0$, $R_i\equiv (R_{0i}, R_{1i})$ is sampled from:
	\begin{equation}
		\begin{split}
			\Pr(R_i=r\mid h_0, h_1, B_i=0, G_i=0)\propto
			&\Phi(h_0(\mathbf{x}_i))^{R_{0i}}(1-\Phi(h_0(\mathbf{x}_i)))^{1-R_{0i}}\times\\
			&\Phi(h_1(\mathbf{x}_i))^{R_{1i}}(1-\Phi(h_1(\mathbf{x}_i)))^{1-R_{1i}}\times \\
			&\ind{r\neq (1,1)},
		\end{split}
	\end{equation}
	which is the joint probability distribution of the latent variables from Eq.~\eqref{eq:Rprobs}, truncated away from the $R_{0i} = R_{1i}=1$  region.\footnote{Formally, this MCMC sampler affects joint updates for $R_i$ and the latent variables in the two probit BART models} For readers interested in convergence properties of our MCMC sampler, we refer you to \autoref{appendix_mcmc}.

\subsection{BART Hyperparameters}
	We run the monotone BART and BART algorithms  with mostly the default specifications of \cite{bart}.  As mentioned in \cite{bart}, a benefit of the BART model is its relative insensitivity to hyperparameter tuning.  Specific specifications that we change relative to the default hyperparameters are that we use 2,000 burn-in draws, 2,000 posterior draws, 1,000 cut-points generated uniformly, and 100 trees (the specifications we use throughout in the simulated data and empirical analysis). For numerical evidence that our methodology performs well under a variety of settings, we refer interested readers to \autoref{sim_study}, where we present results of a robust simulation study.
	
	\color{black}

	\section{Empirical analysis of bankruptcy data}\label{empirical_section}
	
	In this section, we study the question of whether adverse going concern opinions cause bankruptcy. We conduct a modular sensitivity analysis based on a monotone BART model fit. This combination allows us to use machine learning methods to learn potentially complex functional forms for the observable data distribution---while reaping the estimation benefits of imposing monotonicity---and obtain valid measures of uncertainty for average and subgroup average effects under different assumptions about the distribution of private information.
	
	Data collection is described in  \autoref{data_section}.  Results are presented in \autoref{sens_analysis}, specifically posterior summaries of firm-year estimated inducement effects as $f(u)$ is varied.  For illustration, several individual firms are investigated in  \autoref{individ_firms}.  Finally, firm characteristics that moderate the inducement effect are investigated in  \autoref{4.5}.
	\subsection{Data}\label{data_section} Data were collected and merged from Audit Analytics, Compustat, and BankruptcyData.com for the sample period of 2000--2014 leading to 20,773 firm-year observations. Of these, 1,535 received an adverse going concern opinion, 522 filed for bankruptcy the next year, and 282 of these bankruptcies received an adverse going concern opinion the previous year.  The bankruptcy indicator was assigned value of 1 if it occurred within a year of the audit report.  This was done because Statement of Auditing Standards No.~59 requires audit firms to opine whether there is substantial doubt regarding a client's ability to continue operating as a ``going concern'' over the twelve months following the financial statement audit.
	
	The following are the control covariates that constitute $\bm{x}$:
	
	\small 
	\begin{enumerate}\label{covariates}
		\itemsep0em 
		\item {\tt Log(Assets)}: Natural log of total assets
		\item {\tt Leverage}: Ratio of total liabilities to total assets
		\item {\tt Investment}: Ratio of short-term investments to total assets
		\item {\tt Cash}: Ratio of cash and cash equivalents to total assets
		\item {\tt ROA}: Ratio of income before extraordinary items to total assets
		\item {\tt Log(Price)}:  Natural log of stock price
		\item {\tt Intangible assets}: Ratio of intangible assets to total assets
		\item {\tt R\&D}: Ratio of research and development expenditures to sales
		\item {\tt R\&D missing}: Indicator for missing R\&D expenditures
		\item {\tt No S\&P rating}: Indicator for the existence of an S\&P credit rating
		\item {\tt Rating below CCC+}: Indicator for S\&P credit rating below CCC+
		\item {\tt Rating downgrade}: Indicator for an S\&P credit rating downgrade from above CCC+ to CCC+ or below
		\item {\tt Non-audit fees}: Ratio of non-audit fees to total audit fees
		\item {\tt  Non-audit fees missing}: Indicator for missing non-audit fees
		\item {\tt Years client}: Number of years client used auditor
		\item {\tt Average short interest}: Interest expense/total assets
		\item {\tt Short interest ratio}: Average short interest (measured in number of shares)/total shares outstanding three months prior to the auditor signature date
		\item {\tt Sum of log returns}: The sum of log daily return in year $t$
		\item {\tt Return Volatility}: The standard deviation of daily returns in year $t$
		\item {\tt Time fixed effect}: A dummy variable for the years 2000--2014
	\end{enumerate}
	\normalsize
	These variables are similar to those used in \cite{paper}, which were inspired by \cite{defond-2002}, and were chosen due to their potential relevance to a companies' upcoming bankruptcy risk as well as their relevance to the issuance of a going concern opinion. 
	\color{black}
	\subsection{Sensitivity to the distribution of private information}\label{sens_analysis}
	
	For fixed conditional probabilities on $(B, G)$ outcomes in \autoref{long}, different choices of $f(u)$ will yield different causal estimates based on solutions to $(b_0, b_1, g)$. Specifically, the right tail of the density $f(u)$ governs how likely an auditor is to observe information that would make a bankruptcy much more likely than suggested by the available covariates, while the left tail governs how likely an auditor is to observe information that would make bankruptcy much less likely than indicated by the available covariates. For reference, in a bivariate probit analysis, $f(u)$ is assumed to have a $\N(0, \sigma)$ distribution, where $\sigma = \sqrt{\rho/(1-\rho)}$; larger $\sigma$ means the available covariates are a more incomplete guide to actual bankruptcy risk. \autoref{resultssummary_rr} reports estimated inducement effects for various specifications of the standard deviation of the private information, $\sigma = \sqrt{V(U)}$.  \autoref{resultssummary_rr} confirms our intuition that a larger variance on $f(u)$ will shrink both our inducement and risk ratio estimates to a null effect.  From an empirical perspective, the table gives us reasonable confidence that there is indeed an effect of going concern on bankruptcy.  
	\color{black}
	
\begin{figure}[!httb]
		\centering 
		\begin{subfigure}{.5\textwidth}
			\includegraphics[width=7.cm]{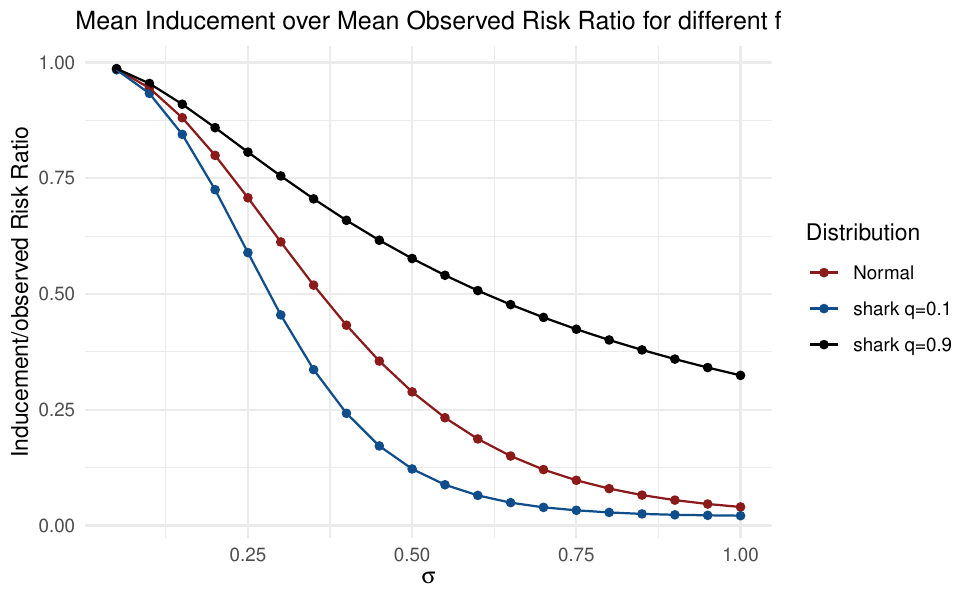}
		\end{subfigure}%
		\begin{subfigure}{.5\textwidth}
			\includegraphics[width=7.cm]{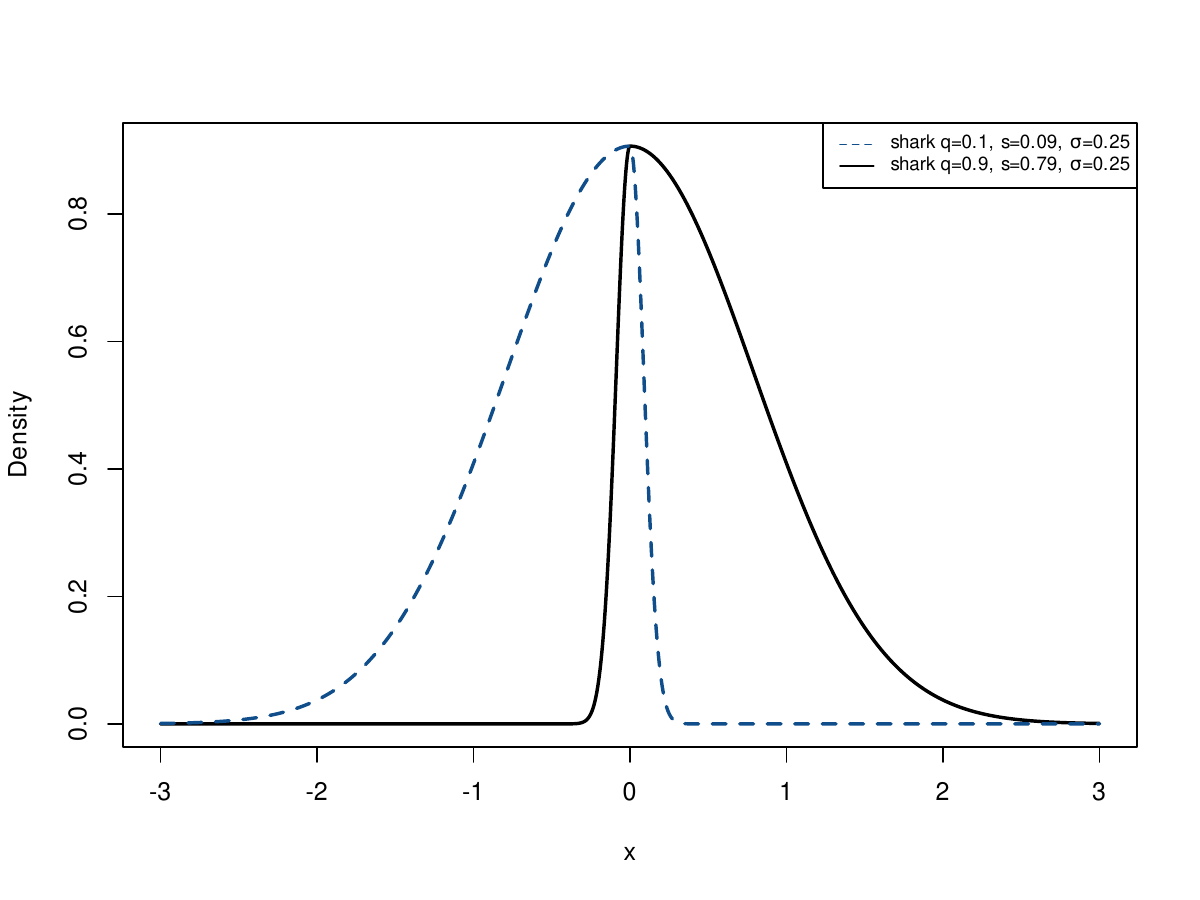}
		\end{subfigure}
		\caption{Left: Plot of inducement effect over observed risk ratio for different standard deviations $\sigma$ for $f$ normally distributed (red), sharkfin ($q=0.9$) with right skew (black), and sharkfin ($q=0.1$) with left skew (blue). The mean observed risk ratio was 52.68. On right is a plot of the shark fin with $q=0.1$ and $q=0.9$, for visual purposes. }
		\label{f_explain_plots}
	\end{figure}

	In addition to varying $\sigma$ for a Gaussian distribution over $U$, we also consider unimodal asymmetric specifications, reflecting the belief that the unreported information is more likely to inflate (or deflate) bankruptcy probabilities even though it is most likely that there is no private information. Specifically, we consider a skewed unimodal (at zero) density with Gaussian tails called the ``sharkfin'' \citep{hahnslice}, which has the following expression:
	\begin{equation}
		\pi(\beta)=\begin{cases}
			2qf(\beta)&\beta\leq 0\\
			2f\qty(\frac{\beta}{1-q}\cdot q)\cdot q&\beta>0
		\end{cases}
		\label{shark}
	\end{equation}
	where $f(\cdot)$ is the pdf of the normal distribution with standard deviation $s$, and $q = \mbox{Pr}(U < 0)$ controls the skewness. The right panel of \autoref{f_explain_plots} depicts two sharkfin densities with $q = 0.1$ and $q = 0.9$ for illustration.
	
	Additionally, we consider two three-component Gaussian mixtures, one symmetric about zero and the other asymmetric with a high weight on the component with the positive mean parameter:
	$$f(u) = 0.05 \phi(u ; -2, s^2) + 0.90 \phi(u ; 0, s^2) + 0.05 \phi(u ; 2, s^2)$$
	and
	$$f(u) = 0.01 \phi(u ; -2, s^2) + 0.94 \phi(u ; 0, s^2) + 0.05 \phi(u ; 2, s^2)$$
	respectively, with $s = 0.05$. Each of these models reflects the case of a small possibility of quite strong positive or negative private information regarding a firm's bankruptcy risk.
	
	\autoref{resultssummary_rr} reports posterior estimates of the average inducement effect across the firms in our study for various choices of $f(u)$. The left panel of \autoref{f_explain_plots} shows the sample average inducement effect (causal risk ratio) as a fraction of the observed risk ratio plotted against $\sigma$ (the standard deviation of $U$) for various  specifications of $f(u)$; consistent with intuition, it shows that greater dispersion of $f(u)$ drives the estimated inducement effect to zero, while the skewness dictates the rate of decay.
	
	\begin{table}[t]
		\centering
		\scalebox{.87}{
		\begin{tabular}{lP{1.9cm}P{2.9cm}P{2.7cm}P{3.1cm}}
			
				\toprule
			Distribution of $f(u)$   &Inducement posterior mean &95\% Credible interval for mean inducement &Risk difference posterior mean& 95 \% Credible interval for mean risk difference \\ \midrule
		$\N(0,\sigma=0.1)$ & 111 & $\qty(39.6,279)$ & 0.100 &  $\qty(0.071, 0.129)$ \\ 
		$\N(0,\sigma=0.5)$& 33.9 & $\qty(11.8 , 91.7)$ & 0.041 &$\qty(0.027, 0.056)$ \\ 
	$\N(0,\sigma=1)$& 4.08 & $\qty(1.82,9.79)$ & 0.007& $\qty(0.004, 0.011)$ \\ 
Shark $q=0.25$, $s=0.5$ ($\sigma=1.05$)& 1.51 & $\qty(1.14, 2.46)$ & 0.003 & $\qty(0.001 , 0.004)$ \\ 
			Shark $q=0.75$, $s=1.25$ ($\sigma=0.88$) 	& 27.8 & $\qty(9.69, 74.9)$ & 0.028 &$\qty(0.018, 0.040)$ \\ 
			
			Symmetric mixture ($\sigma=0.64$)&24.4 & $\qty(8.40 , 64.2)$ & 0.023 & $\qty(0.019  , 0.029)$ \\ 
			Asymmetric mixture ($\sigma=0.49$)& 25.6 & $\qty(8.09 , 72.0)$ & 0.025 & $\qty(0.018 , 0.031)$  \\

			\bottomrule

		\end{tabular}
	}
		\caption{Posterior mean estimates and credible intervals for inducement effect and risk differences. See \autoref{sec:RDvsInd} for a discussion on differences vs. inducement.  The reduced form probabilities in \autoref{long} were estimated using BART with a monotonicity constraint on the going concern variable.  We further require $b_1(\bm{x})>b_0(\bm{x})$ in the projection step. Posterior summaries based on 500 Monte Carlo samples (of the posterior draws for all the firms in our dataset). $\sigma$ refers to the implied standard deviations of the different distributions. }
		\label{resultssummary_rr}
	\end{table}
	\color{black}
	
	\subsection{Comparison with the E-value}\label{E-value-comparison}
\begin{figure}[!httb]
		\centering
		\begin{minipage}{.6\textwidth}
			
			\begin{tikzpicture}[thick,scale=.65, every node/.style={scale=.65}]
				\node (X) at (0,0) {$\boxed{\text{Going concern}}$};
				\node at (0,.5) {$G$};
				\node (Y) at (-4,0)  {Observed Covariates};
				\node at (-4,-.5) {$\bm{x}$};
				\node (Z) at (3,2) {Unobserved};
				\node at (3,2.5){$U$};
				\node (U) at (6,0)  {$\boxed{\text{Bankruptcy}}$};
				\node at (6,.5) {$B$};
				\draw[thick, -Latex](X)--(U);
				\draw [thick,-Latex] (Y) -- (X);
				\draw [thick, dashed, >=triangle 45, <->] (Y) to[bend right=-30] (Z);
				\draw [thick,-Latex] (Y) to[bend right=30] (U);
				\draw [thick,-Latex] (Z) -- (X) node [midway,below,sloped] {$\text{RR}_{GU}$};
				\draw[thick,-Latex] (Z) -- (U)  node [midway,below,sloped] {$\text{RR}_{UB}$};
			\end{tikzpicture}
			
		\end{minipage}%

		\caption{$\text{RR}_{GU}$ is the maximum risk ratio comparing any two categories of confounding and $\text{RR}_{UB}$ is the
			maximum risk ratio for any specific level of the unmeasured confounders comparing those with and without treatment, controlling for $\bm{x}$.}
		\label{rel_risk_table}
	\end{figure}
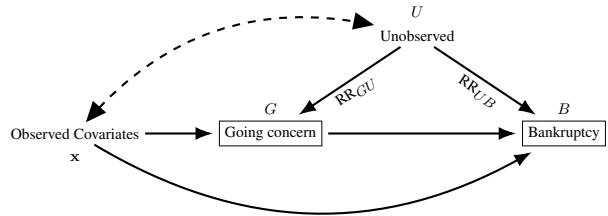

	Rather than modeling the distribution of unobserved information $f(u)$, an alternative approach is to consider the strength of unobserved confounding that would be necessary to entirely explain the observed association. This approach can be found as early as \cite{Cornfield}, and has recently been generalized in \cite{evalue} and \cite{Peng-2016}, who prove that 
	
	\begin{equation}
		\max(\text{RR}_{GU}, \text{RR}_{UB})\geq \cbr{\text{RR}^{\text{obs}}_{GB}+\sqrt{\text{RR}^{\text{obs}}_{GB}\qty(\text{RR}^{\text{obs}}_{GB}-\text{RR}^{\text{true}}_{GB})}}\bigg{/}\text{RR}^{\text{true}}_{GB}
		\label{high_thresh}
	\end{equation}
	where
	\begin{align*}
		\text{RR}_{GU\mid \bm{x}} &= \max_{k} \;\;  \frac{\Pr\qty(U=k\mid G=1, \bm{x})}{\Pr\qty(U=k\mid G=0, \bm{x})},\\
		\text{RR}_{UB\mid \bm{x}} &= \max_{k,k',g} \;\; \frac{\Pr\qty(B=1\mid G=g, \bm{x}, U=k)}{\Pr\qty(B=1\mid G=g, \bm{x}, U=k')} 
	\end{align*}
	for $g \in \lbrace 0,1\rbrace$ and
	\begin{equation}
		\text{RR}^{\text{true}}_{GB}=\frac{\int \Pr(B=1\mid G=1, \bm{x}, U)\Pr(U\mid \bm{x})\dd u}{\int \Pr(B=1\mid G=0, \bm{x}, U)\Pr(U\mid \bm{x})\dd u}.
		\label{true_causal_rr}
	\end{equation} 
	\autoref{rel_risk_table} provides a visualization of these terms.
	
\begin{figure}[h]
		\centering
		\includegraphics[height=8.25cm]{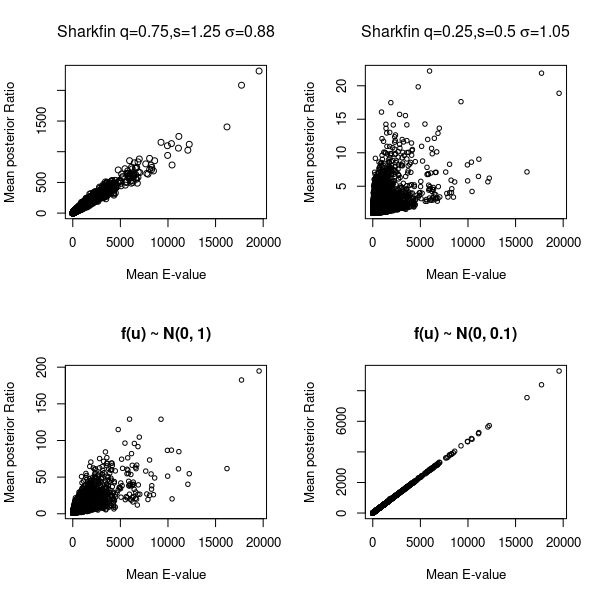}

		\caption[E-value vs ratio]{  Posterior means of $\tau$ across 500 draws for different distributions of $f(u)$ vs. the E-value per firm calculated from the posterior mean of the risk ratio from $\text{RR}^{\text{obs}}_{GB\mid \bm{x}}=\Pr\qty(B=1\mid G=1, \bm{x})/\Pr\qty(B=1\mid G=0, \bm{x})$. }
		\label{E-val-ratio_audit}
		
	\end{figure}

	Setting $\text{RR}_{GB}^{\text{true}}=1$ in expression \autoref{high_thresh}, \cite{Peng-2016} define the {\em E-value} (for evidence value) as
	\begin{equation}m
		\text{E-value} = \text{RR}^{\text{obs}}_{GB}+\sqrt{\text{RR}^{\text{obs}}_{GB}\qty(\text{RR}^{\text{obs}}_{GB}-1)},
		\label{eval_full}
	\end{equation} 
	which can be interpreted as the minimum strength of association that an unmeasured confounder would need to have with both $G$ and $B$ (conditional on $\bm{x}$) to fully explain the observed treatment-outcome association. Note that for large observed risk ratios (that is, $RR^{\text{obs}} \approx RR^{\text{obs}} - 1$), the E-value is essentially proportional to the observed risk ratio itself. Accordingly, if we compare our model-based sensitivity analysis estimates to the E-value, we find that when $f(u)$ concentrates around zero, the associated causal risk ratio becomes the observed risk ratio, which is effectively the E-value. However, for different choices of $f(u)$, the associated causal risk ratio at different $\bm{x}$ values can differ from the observed risk ratio in interesting ways, which we explore in the following sections. \autoref{E-val-ratio_audit} plots posterior means of $\tau$ against the posterior mean of the E-value for the auditing data for the distributions of $U$ reported in \autoref{E-val-ratio}. Essentially, E-values are simply a scale multiple of the observed risk ratio, which is precisely the causal risk ratio when there is assumed to be no private information (lower right panel of \autoref{E-val-ratio_audit}). However, less dogmatic choices of $f(u)$ also yield substantial inducement effect estimates for some firms (first three panels of \autoref{E-val-ratio_audit}).
	
	\subsection{Posterior Individual Inducement Effects for Specific Firms}\label{individ_firms}
\begin{figure}[!httb]
		\centering 
			\begin{minipage}{.3\textwidth}
			\includegraphics[width=4.5cm]{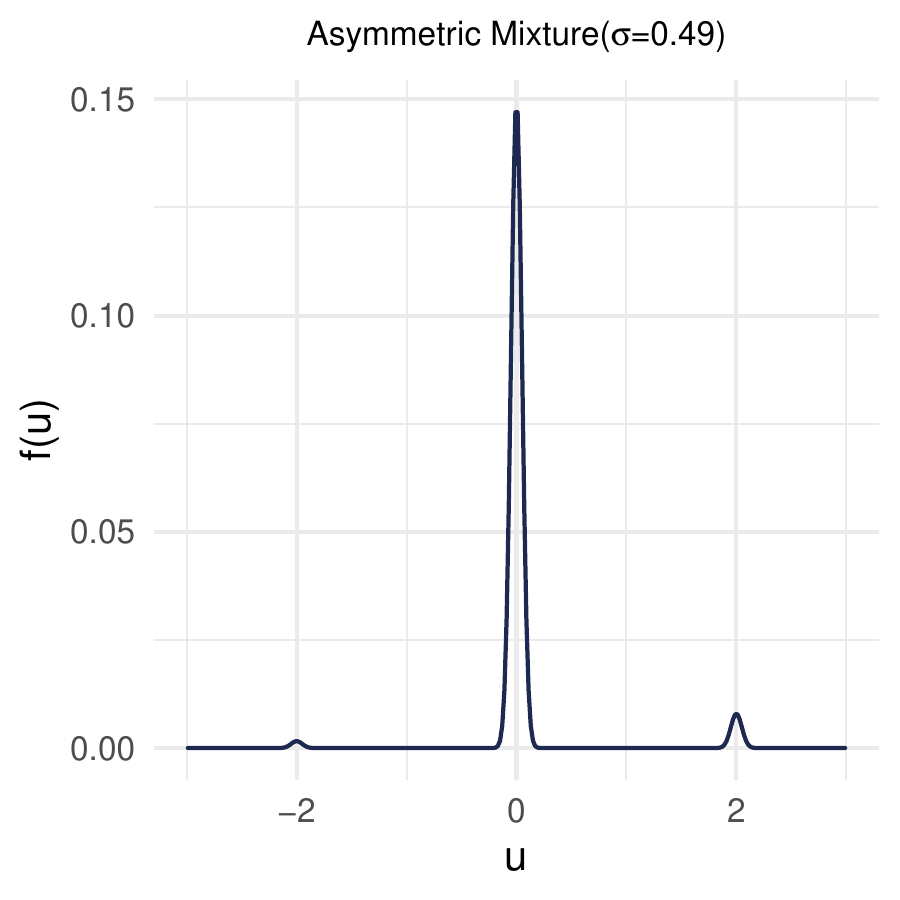}
		\end{minipage}%
		\begin{minipage}{.5\textwidth}
			\includegraphics[width=9cm]{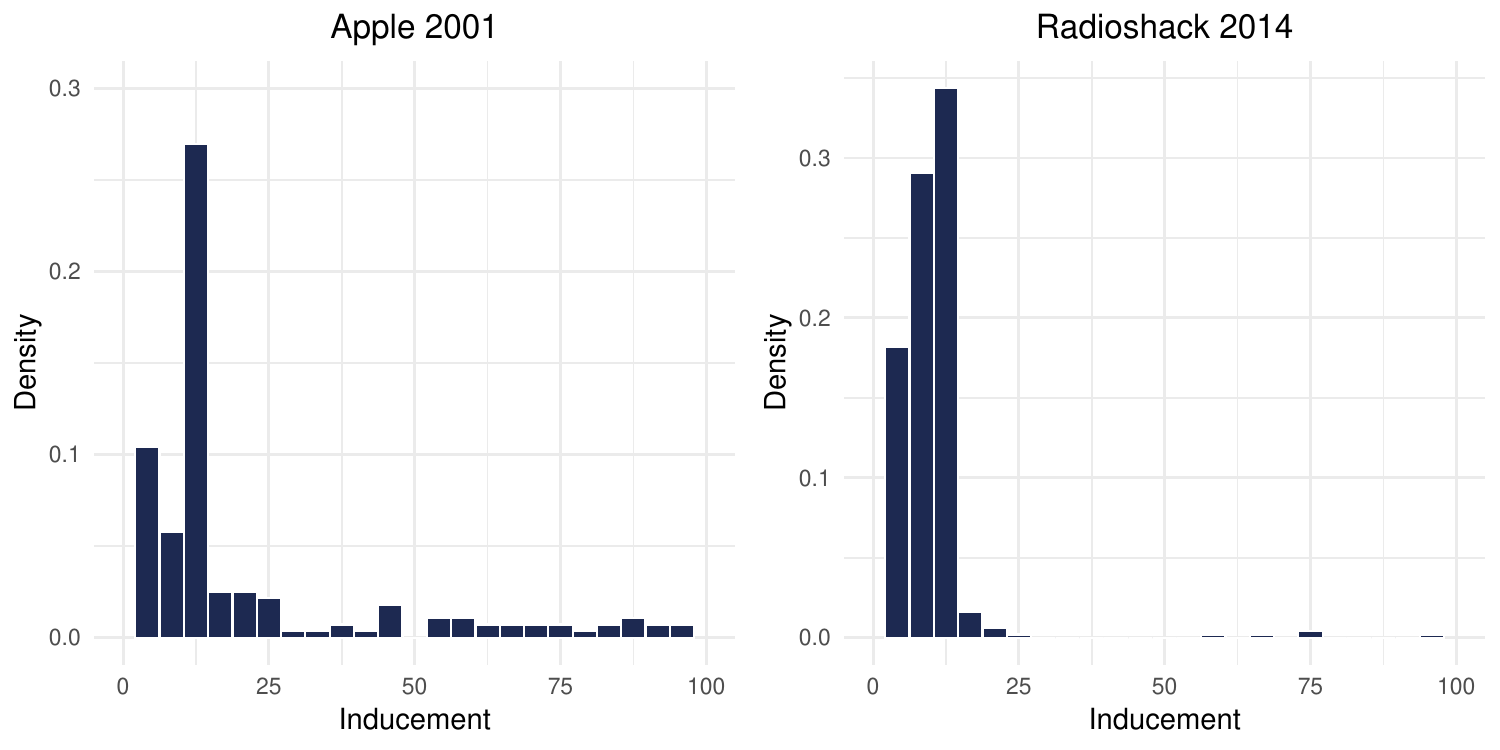}
		\end{minipage}
	\caption[Apple vs Radioshack]{Histogram of posterior estimates of the individual inducement effects given $f(u)$ with the distribution on the left (moderate confounding) for Apple 2001 and Radioshack 2014.  Neither received an adverse going concern, but Radioshack did go bankrupt. Radioshack was audited by PWC, and Apple was audited by KPMG.}

		\label{individ_firm_plot}
	\end{figure}

	By numerically solving \autoref{long} for $(b_0, b_1, g)$ at each posterior draw, for a given firm-year observation and a given choice of $f(u)$, a full posterior distribution over causal estimands for that observation can be obtained.  Scrutinizing these posteriors for specific firms provides an intuitive approach to investigating the results of the sensitivity analysis that is more granular than simply reporting sample averages across all observations. To this end, the posterior mean inducement effect, as well as a 95\% credible interval, are presented in \autoref{individ_firm_table} for a selection of illustrative firms.  \autoref{individ_firm_plot} depicts a histogram of posterior draws of the inducement effect for Apple (from year 2001) and Radioshack (from 2014) using an asymmetric Gaussian mixture.

	\begin{table}[!httb]
	\centering
	
	\begin{subtable}{.71\textwidth}
		\scalebox{0.65}{
			\begin{tabular}{lP{1.1cm}P{1.1cm}P{1.1cm}P{1.1cm}P{1.1cm}P{1.cm}P{1.cm}P{1.8cm}}
				\toprule
				Firm   &Going Concern&Bankruptcy& Auditor&mean $\text{RR}_{\text{obs}}$&mean $B_0$& mean $B_1$&mean $\tau$ post &95\% Credible interval for $\tau$ (\%)  \\ \midrule
				JetBlue (2007)&No&No&E\&Y&4.96&0.005&0.062&1.96&$(1.00, 6.16)$\\
			JetBlue (2009)&No&No&E\&Y&44.1&0.001&0.011&12.6&$(1.19, 51.8)$\\
				Apple (2001) &No&No&KPMG&957&0.001&0.024&247&$(11.2, 1392)$\\
				Build a Bear (2010)&No&No&KPMG&177&0.001&0.021&57.1&$(3.21, 463)$\\
				Build a Bear (2014)&No&No&E\&Y&18.1&0.005&0.030&6.93&$(1.45, 21.3)$\\
				Radioshack (2014)&No&Yes&PWC&51.4&0.002&0.015&14.3&$(1.49, 59.2)$\\
				Blockbuster (2004)&No&No&PWC&48.1&0.004&0.035&17.0&$(1.48, 76.6)$\\
				Blockbuster (2009)&Yes&No&PWC&7.99&0.029&0.107&4.21&$(1.51, 10.8)$\\
				Six Flags (2006)&No&No&KPMG&12.8&0.010&0.046&5.84&$(1.31, 18.9)$\\
				Six Flags (2009)&Yes&Yes&KPMG&3.33&0.037&0.052&1.71&$(1.00, 5.14)$\\ \midrule
				Largest RD Sub&188&41&&19.3&0.073&0.193&6.24&$\qty(1.19, 28.0)$\\
				Largest RR Sub&178&52&&540&0.002&0.041&63.2&$\qty(3.91, 234)$\\
				\bottomrule          
			\end{tabular}
		}
	\end{subtable}%
	\vline\vline
	\begin{subtable}{.71\textwidth}
		\scalebox{.65}{
			\begin{tabular}{P{1.cm}P{1.cm}P{1.cm}P{1.8cm}}
				
				\toprule
				mean $B_0$& mean $B_1$&mean $\tau$ post &95\% Credible interval for $\tau$ (\%)  \\ \midrule
				0.030&0.035&1.20&$(1.00, 2.64)$\\
				0.005&0.024&8.26&$(1.08, 73.3)$\\
				0.002&0.034 &284&$(1.05, 2159)$\\
				0.005&0.021&42.2&$(1.09, 518)$\\
				0.008&0.011&1.42&$(1.00, 2.92)$\\
				0.005&0.034&11.1&$(1.08, 70.8)$\\
				0.008&0.023&5.51&$(1.07, 11.0)$\\
				0.040&0.063&1.64&$(1.00. 4.19)$\\
				0.014&0.018&1.32&$(1.00, 3.34)$\\
				0.044&0.047&1.09&$(1.00, 2.06)$\\ \midrule 
					0.075&0.151&2.42&$\qty(1.05, 6.60)$\\
				0.005&0.036&15.9&$\qty(1.69, 67.9)$\\
				\bottomrule       
			\end{tabular}
		}
	\end{subtable}
	\caption[Different ACRR estimates for specific firms]{Left: Posterior estimates of the inducement  effect given $f(u)\sim \N(0, \sigma=0.5)$ for select firms.  Right: Posterior estimates of the inducement  effect given $f(u)$ is the asymmetric Gaussian mixture with an upweighted right component for the same firms. The second from bottom row references the bottom right of the tree (left panel) of \autoref{cart_tree_treat}, which are the firms with the largest subgroup risk difference effects. The bottom row is referencing the bottom right of the tree (the left panel) of \autoref{cart_tree_RR}, which are the firms with the largest subgroup risk ratio effects. The going concern and bankruptcy columns in these two rows refer to the number of firms in those respective subgroups that were issued going concern opinions and filed for bankruptcy.  }
	\label{individ_firm_table}

\end{table}

	We find that the inducement effect varies both across posterior draws as well as across firms as a function of the density $f(u)$.  Differences between firms are illuminating: for example, Apple in 2001 had a significantly higher inducement effect than Blockbuster in 2009, but this is at least in part an artifact of Apple 2001 having an extremely low probability of bankruptcy. This points to a general phenomenon with risk ratios, which is that they can be dramatically impacted by the denominator; we explore this fact further in the following section.
	
	\subsection{Exploratory subgroup analysis}\label{4.5} 

	With firm-year specific treatment effects in hand, one can conduct an ex post regression tree analysis to isolate subgroups of firms with subgroup average treatment effects that depart from the overall average.
	Specifically, we identify moderating subgroup of variables by fitting a single regression tree using the individual inducement effect estimates (posterior means) as the response variable and observable firm (and auditor) features as predictors (as detailed in \cite{bcffreak}).  For predictors we use the same covariates reported in \autoref{covariates}, all of which are plausible moderators of the inducement effect.

	\begin{figure}[t]
		\centering
		\begin{subfigure}{.5\textwidth}
				\includegraphics[ width=7cm]{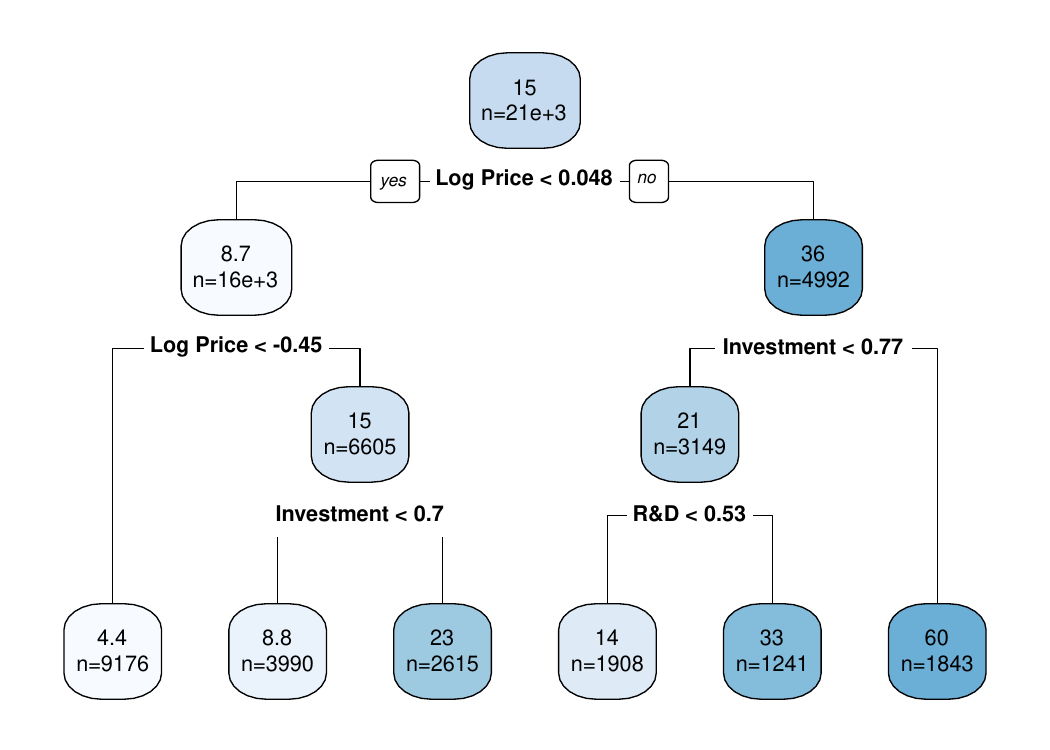}

		\end{subfigure}%
		\begin{subfigure}{.5\textwidth}
			\includegraphics[ width=8cm]{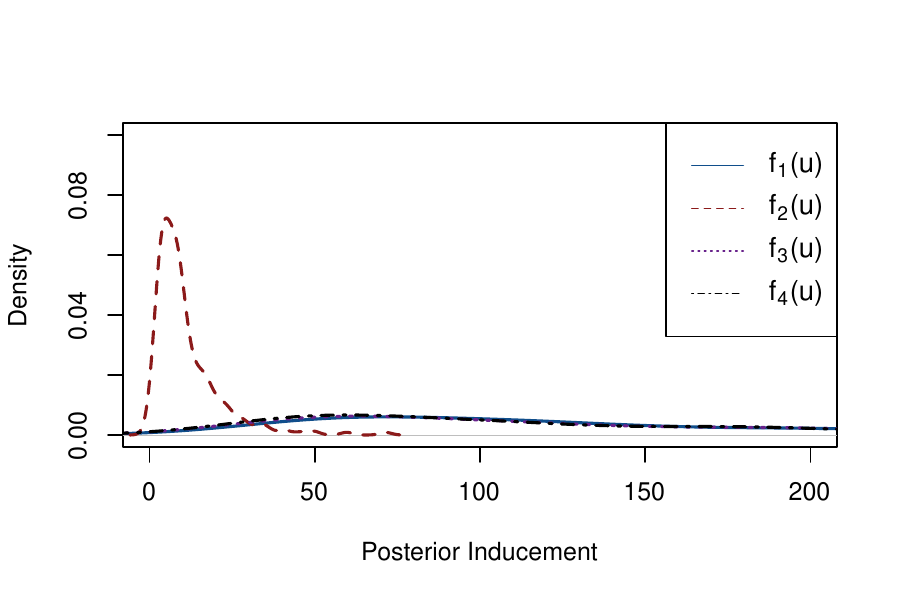}

		\end{subfigure}
		\caption{Left: A small tree fit to inducement effects (risk ratios).  This is also the group of variables we investigate as moderators. Follow down tree to identify subgroup.  Right: Plot of difference of inducement effects across the posterior draws between the largest and smallest inducement effect subgroups (bottom right and bottom left respectively on the tree). }
		\label{cart_tree_RR}
	\end{figure}

	The subgroup analysis presented here is based on $U \sim \N(0, \sigma=0.5)$ to the left hand side of \autoref{long}. The left panel of \autoref{cart_tree_RR} shows the resulting tree fit. Using this tree, we can identify subgroups based on the corresponding partition implied by terminal node (leaf) membership. However, the resulting point estimates only tell part of the story. For a fuller picture, we can consider the posterior distribution of subgroup {\em differences}, even for different choices of $f(u)$ than the one used to produce the tree.  We compute the subgroup difference of mean inducement effects for each posterior draw between the subgroups with the largest and smallest subgroup effects as determined by the regression tree.  This analysis is repeated for four different distributions of $f(u)$: $f_1(u)\sim \N(0,\sigma=0.5)$, $f_2\sim \N(0,1)$, $f_3(u)$ which is a mixture model with more weight on a far bump to the right (see \autoref{individ_firm_plot}), and $f_4(u)$ which is a three component Gaussian mixture with 90\% of the area centered around 0, and 5\% around $u=-2$ and $u=2$. The right panel of \autoref{cart_tree_RR} shows posteriors of subgroup differences in inducement effects (causal risk ratios); the sign of the differences is preserved across various choice, while the magnitude varies (as one might anticipate).
	
	With respect to economic interpretation, the tree presented in \autoref{cart_tree_RR} shows that firms with higher stock prices [{\tt Log(Price)}] and greater investments [{\tt Investments} and {\tt R\&D}] have higher risk ratios. Prior studies find that firms with higher stock prices \citep{Campbell-Hilscher-Szilagyi-2008} and greater R\&D have lower bankruptcy risk \citep{JindalMcAlister-2015}. Firms with greater fixed assets investments are also considered being relatively ``safe.'' Thus, the higher risk ratios for these firms are likely driven by small denominators. 
	
	\subsubsection{Risk Difference vs. Inducement}\label{sec:RDvsInd}
\begin{figure}[t]
		\centering
		\begin{subfigure}{.5\textwidth}
				\includegraphics[width=7cm]{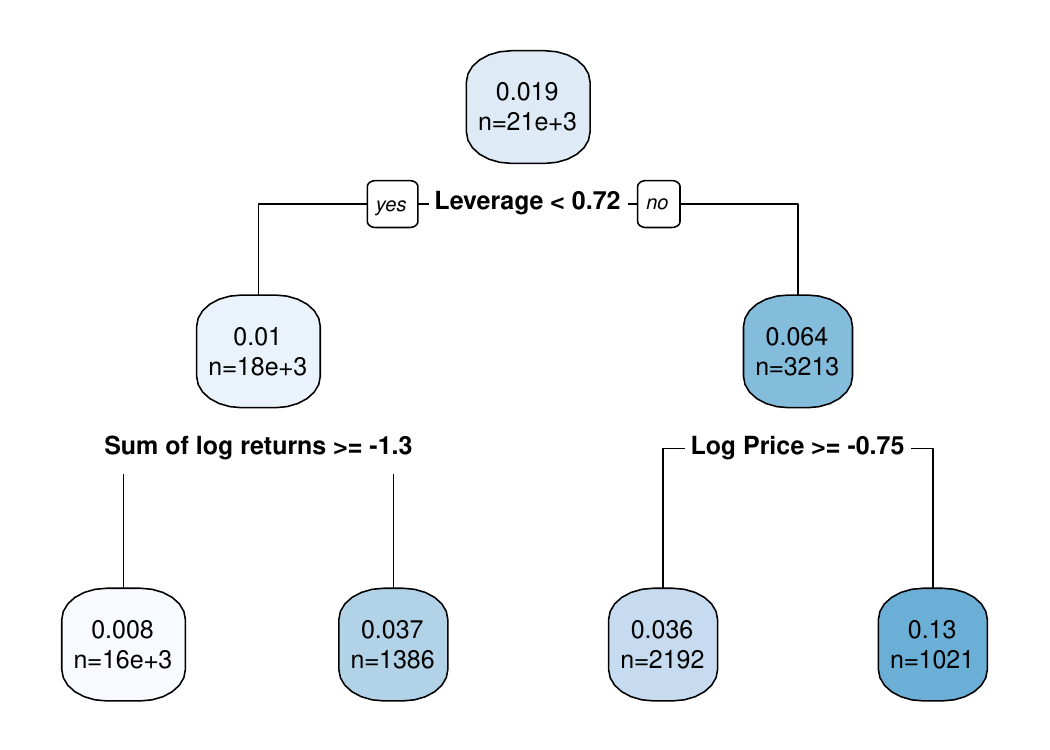}
		\end{subfigure}%
		\begin{subfigure}{.5\textwidth}
				\includegraphics[ width=8cm]{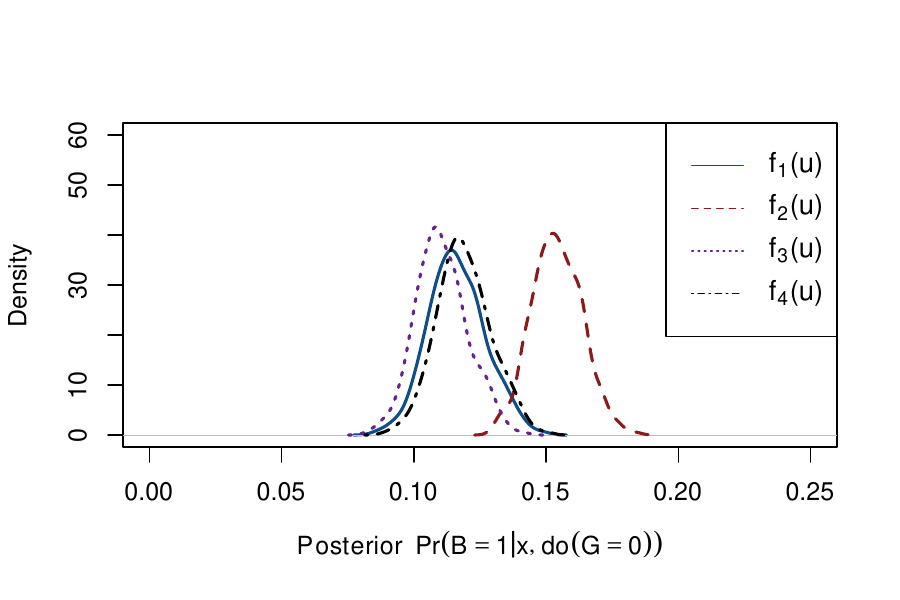}
		\end{subfigure}
		\caption{Left: A small tree fit to the $\Pr(B=1\mid \bm{x}, \text{do}(G=0))$, $B_0$ for short.  Right: Plot of differences of $B_0$  across the posterior draws between the largest and smallest $B_0$ effect subgroups (bottom right and bottom left respectively on the tree).}
		\label{B0_tree_fig}
	\end{figure}

	At this point, it is instructive to consider whether different estimands may be moderated by different covariates. In particular, the results in \autoref{cart_tree_RR} suggest that risk ratios may be dominated by the denominator, which may be affected by different variables than those which affect the numerator. Accordingly, in \autoref{B0_tree_fig}, we fit a regression tree to point estimates of the $\Pr(B=1\mid \bm{x}, \text{do}(G=0))$. For this tree, we find that firms with higher leverage and lower stock returns and prices have higher probabilities of bankruptcy absent a going concern opinion.\footnote{\cite{Campbell-Hilscher-Szilagyi-2008} find similarly that the probability of bankruptcy increases in leverage and decreases in share price.}
	
	\color{black}
	
	We next consider the risk difference $\Pr(B =1\mid \bm{x}, \text{do}(G=1))-\Pr(B=1\mid \bm{x}, \text{do}(G=0))$. While risk ratios can be unappealingly large for firms with very small bankruptcy risk, risk differences (necessarily) have the opposite complication, which is that a difference of 0.1 ``means'' something quite different for a firm with control probability of 0.5 than it does for one with control probability 0.9. Fortunately, the risk difference has another interpretation in contexts like the present one where treatment effects are assumed to be monotonic: the risk difference is equivalent to the probability that a firm went bankrupt {\em because of} the going concern opinion. This interpretation is derived as follows. Consider the four possible potential outcomes, depicted in \autoref{tab:induce_explain_table}, which gives each configuration a suggestive name.
	
	\begin{table}[h]
		\centering
		\begin{tabular}{lP{1cm}P{1.5cm}}
			
			\toprule
			Name&$B^{1}$&$B^{0}$\\ \midrule
			No Inducement&0&0\\
			Prevention&0&1\\
			Induced Bankruptcy&1&0\\
			No Prevention&1&1\\
			
			\bottomrule
		\end{tabular}
		\caption{Because we are operating in the binary treatment/binary response world, we have just four outcomes.  The first row refers to a firm that, regardless of a receiving going concern opinion, does not go bankrupt (``No Inducement'').  ``Prevention'' refers to the situation in which, without the treatment, the firm would have gone bankrupt, but with the going concern opinion it does not. We do not allow for this situation given our monotonicity assumption $\Pr(B=1\mid \bm{x}, G=1)\geq \Pr(B=1\mid \bm{x}, G=0)$. ``Induced bankruptcy'' refers to the situation in which the firm goes bankrupt because of the going concern opinion. ``No prevention'' means, regardless of a going concern opinion being issued, the company goes bankrupt.}
		\label{tab:induce_explain_table}
	\end{table}
	
	The marginal probabilities are then simply the sum of rows where ``1'' appears in the corresponding column of \autoref{tab:induce_explain_table}:
	\begin{equation}
		\begin{split}
			\Pr(B=1\mid \bm{x}, \text{do}(G=0))&=\Pr(\text{Prevention})+\Pr(\text{No prevention})\\
			\Pr(B=1\mid \bm{x}, \text{do}(G=1))&=\Pr(\text{Induced bankruptcy})+\Pr(\text{No prevention})
		\end{split}
	\end{equation}
	But, under the monotonicity assumption, $\Pr(\text{Prevention}) = 0$, in which case
	\begin{equation}
		\Pr(B=1\mid \bm{x}, \text{do}(G=1))-\Pr(B=1\mid \bm{x}, \text{do}(G=0))=\Pr(\text{Induced  bankruptcy}).
		\label{prevent_eq}
	\end{equation}
	
\begin{figure}[t]
		\centering
		\begin{subfigure}{.5\textwidth}
				\includegraphics[width=7cm]{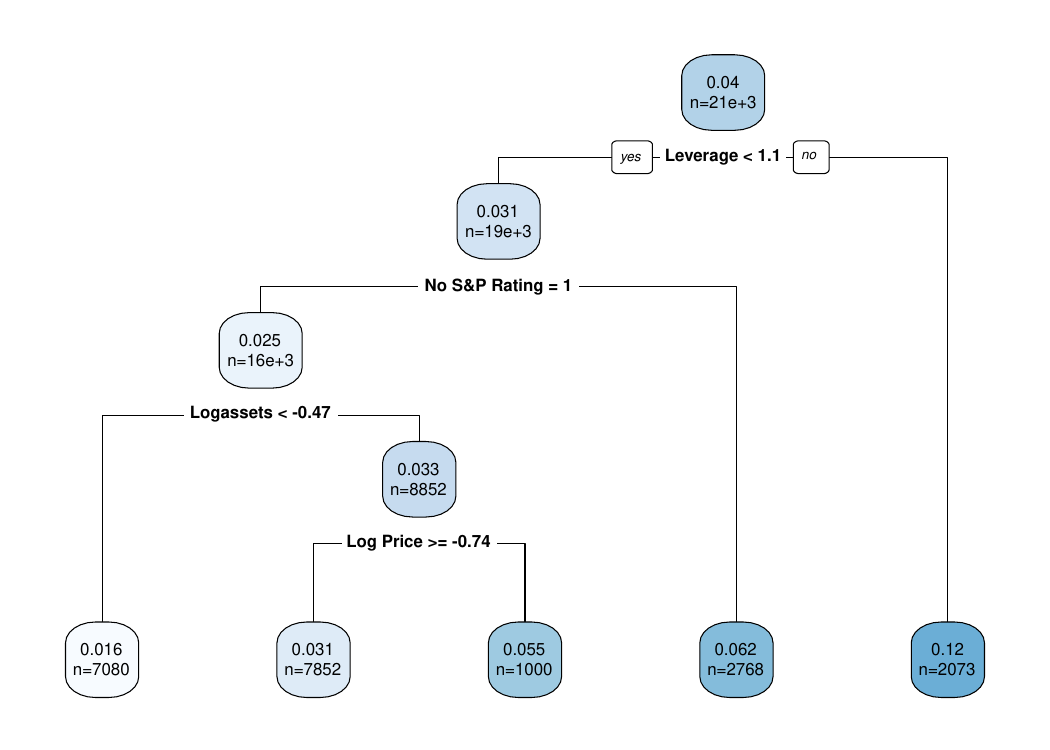}
		\end{subfigure}%
		\begin{subfigure}{.5\textwidth}
				\includegraphics[ width=8cm]{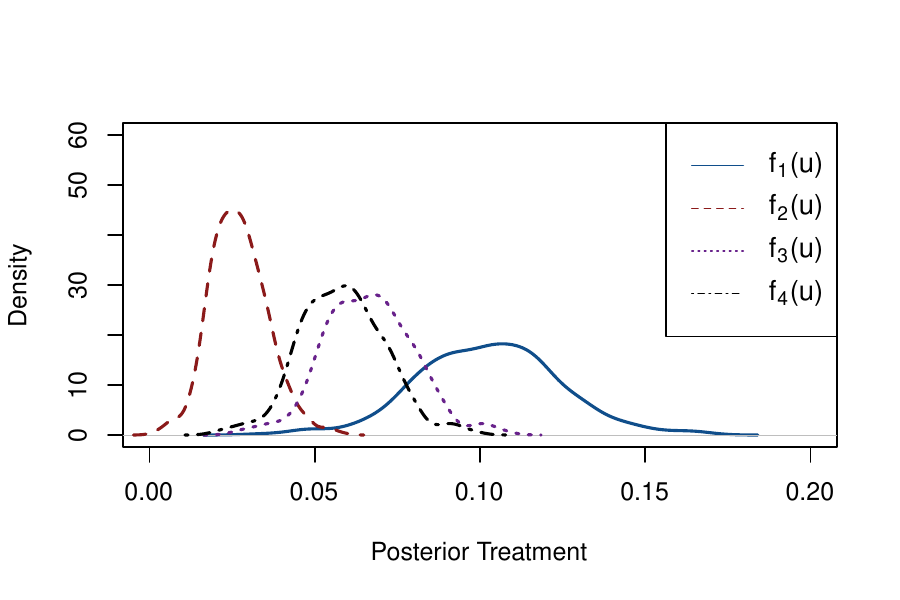}
		\end{subfigure}
		\caption{Left: A small tree fit to the risk difference, $\Pr(B =1\mid \bm{x}, \text{do}(G=1))-\Pr(B=1\mid \bm{x}, \text{do}(G=0))$, which under monotonicity of going concerns, is equivalent to the probability that the bankruptcy was induced.  Right: Posterior subgroup differences between the largest and smallest treatment effect subgroups (bottom right and bottom left respectively on the tree).}
		\label{cart_tree_treat}
	\end{figure}

	Accordingly, in \autoref{cart_tree_treat}, we fit a regression tree to point estimates of the (causal) risk difference $\Pr(B =1\mid \bm{x}, \text{do}(G=1))-\Pr(B=1\mid \bm{x}, \text{do}(G=0))$. At the top of the tree, we find that firms with greater leverage are more likely to have an inducement effect. This result is consistent with \cite{Chen-He-Ma-etal-2016}, who find that debt contracts often include covenants that mechanically increase interest rates when the borrow receives an adverse going concern opinion. At the second level, we find inducement is likely to occur when the firm has an S\&P credit rating. Consistent with this result, \cite{Feldmann-Read-2013} find that S\&P tends downgrade credit ratings after the issuance of a going concern opinion. At the third level, larger firms are more likely to have an inducement effect. This result could be due to firms' information environments varying with firm size.

	It bears emphasis that the tree-based posterior subgroup analysis presented above is simply an exploration of the posterior distribution. Consequently, the posterior difference shown in the right panels of \autoref{cart_tree_RR}, \autoref{B0_tree_fig}, and \autoref{cart_tree_treat} require no further adjustment. Similarly, the CART fits presented in the left panels cannot be ``over-fit.'' The posterior distribution is where the inferences are performed, CART is being used merely as a way to navigate a high dimensional posterior. Trees are restricted to be small to ease interpretation and to focus on subgroups with relatively large sample sizes. Ideally, these summaries would not be endpoints of an analysis, but the starting point for further investigation into the moderating role of particular attributes.

	\subsection{Comparison with Bivariate Probit}\label{section:bivar_prob_compare}
\begin{figure}[t]
		\centering
		\includegraphics[width=10cm]{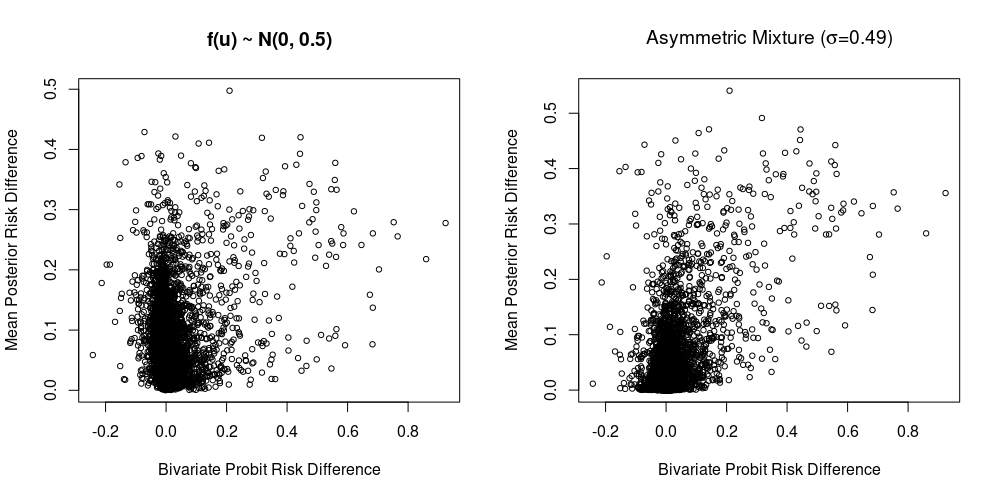}
		\caption{On the horizontal axis is the estimates of the bivariate probit regression with seed \emph{25678}. The estimate of $\rho$ using the bivariate probit regression was $0.420$, which corresponds to a standard deviation (in our model's form) of $\sqrt{\rho\cdot(1-\rho)}=0.494$. On the verical axis is the mean of the posterior risk differences for every firm across two different configurations of $f(u)$.  On the left panel, $f(u)$ is distributed $N(0, 0.5)$, and on the right the distribution is the asymmetric mixture from the left panel of \autoref{individ_firm_plot} ($\sigma=0.486$).}  
		\label{fig:bivar_vs_our}
	\end{figure}

In this section, we compare our methodology directly to an implementation of the bivariate probit regression for the auditing data.  Our estimand of interest is the risk difference in this section (see \autoref{sec:RDvsInd} for a discussion on risk differences vs. inducement), as this estimand more clearly shows the advantages of our flexible model.  \autoref{fig:bivar_vs_our} shows two advantages of our methodology.  First, as discussed throughout the document, we can model a variety of different ``confounding'' situations through the choice of $f(u)$.  Second, our model seemingly gives more reasonable estimates.  Because of our monotonicity restraint, we do not have any negative estimates, whereas the bivariate probit regression included multiple negative estimates of the risk difference, which in the context of the problem does not appear reasonable.  \autoref{fig:bivar_vs_our} specifically compares two $f(u)$ configurations with standard deviations similar to the estimated value from the bivariate probit regression $\rho=0.420$.  As we show in \autoref{bivar_simsec}, if the true data generating process is from a bivariate probit model and our choice of $f(u)$ is distributed $\N(0, \sigma=\sqrt{\rho(1-\rho)})$, both methods return similar results.  If the data generating process is \emph{not} the bivariate probit, our method still recovers true estimates well whereas the bivariate probit does not.  Therefore, \autoref{fig:bivar_vs_our} gives more credence to the theory that the true data generating process is unlikely to be  a bivariate probit model.  
	
	\section{Simulation studies}\label{sim_study}
	In this section we investigate how the new method performs under a variety of different simulated data generating processes and modeling assumptions, in an effort to build confidence in the empirical analysis above. 
	\begin{itemize}
	\item \autoref{bivar_simsec} shows that when the data are generated according to a linear bivariate probit model our new approach is able to recover the the true parameter values, despite being based on a more flexible non-parameteric machine learning specification.	
	\item \autoref{nonlindgp} shows that our approach can recover true average causal risk ratios when the data is generated according to more complicated non-linear data generating processes (and $f$ is correctly specified). This section also explores how misspecification of $f$ affects the accuracy of the treatment effect estimates.    
	\item \autoref{E-val_sec} compares the estimated risk ratios to the corresponding E-values (as was done in \autoref{E-val-ratio_audit}), but using simulated data.
	\item   \autoref{mono_section} demonstrates the improved statistical precision of using the monotonicity constraint in BART.
	\end{itemize}
	
	\subsection{Evaluation on bivariate probit data}\label{bivar_simsec} To verify that the proposed machine learning sensitivity analysis yields sensible answers, we take advantage of the relationship between our model and the bivariate probit model with endogenous binary regressor: if we generate the data from the bivariate probit model with $U \sim \mbox{N}(0, \sigma=\sqrt{\rho/(1-\rho)})$, the true causal risk ratios should be recoverable.\footnote{Note, the success of our sensitivity analysis is predicated upon  minimizing the squared distance between the three left hand-side pairs in \autoref{long}. We use the Nelder-Mead algorithm to do so, a commonly used numerical method for minimization of loss functions \citep{nelder} (although we also employed a simulated annealing approach and the Broyden–Fletcher–Goldfarb–Shanno algorithm, both giving similar results as Nelder-Mead).}  \autoref{bivartable} reports the results of fitting our sensitivity analysis model data generated from the bivariate probit
	\[
	\begin{pmatrix}
		Z_{g,i}\\
		Z_{b,i}
	\end{pmatrix}\stackrel{\text{iid}}{\sim}\mathcal{N}(\boldsymbol{\mu}, \bm{\Sigma})
	\qquad \boldsymbol{\mu}=\begin{pmatrix}
		\beta_0+\beta_1\bm{x}_i\\
		\alpha_0+\alpha_1\bm{x}_i
	\end{pmatrix}
	\qquad
	\bm{\Sigma}=\begin{pmatrix}
		1&\rho\\
		\rho&1
	\end{pmatrix}
	\label{model3}
	\]
	We simulated 25,000 samples, where we sum five uniform($-$1,1) $\bm{x}_i$ covariates each with the same $\beta_1$ and $\alpha_1$ coefficients respectively.  We set $\beta_0=0, \beta_1=-0.2, \alpha_0=-0.5, \alpha_1=-0.5$ to generate a reasonable number of going concerns and bankruptcies.  We fit the left hand side of \autoref{long} using BART with the monotonicity constraint, whose benefit is shown in \autoref{mono_section}.  Note, in our simulations, we do not solve our systems for every BART posterior estimate of \autoref{long} due to computational constraints.  Instead, we take the mean of the posterior BART probability estimates in the fitting stage and then solving for our causal parameters $b_0(\vdot), b_1(\vdot), g(\vdot)$ once for each observation.\footnote{This methodology held for all the simulated data; when analyzing the real data we repeated the integrals for random samples of the posterior BART estimates.} We impose the constraint that $b_1(\bm{x})\geq b_0(\bm{x})$ when solving for the causal parameters.

	In simulations (included in \autoref{machine_append}), we observe that at $N=100,000$ the bivariate probit regression (unsurprisingly) works remarkably well when the data generation process is in fact a bivariate probit model.\footnote{It is well-known that maximum likelihood estimates of the bivariate probit model can be unstable (i.e., many local modes), especially when there are a large number of predictor variables (see \cite{Meng-Schmidt-1985} and \cite{Freedman-Sekhon-2010}). Our simulations bear this out; with thousands of observations, estimates of $\rho$ were quite inaccurate. Therefore, to verify that we obtain consistent parameter estimates with maximum likelihood estimation (and to cross-check our data generating process), we  generate and train our models on 100,000 observations (see the supplementary material \citep{audit-supplement}).} At lower sample sizes, maximum likelihood estimation of the bivariate probit with endogenous regressor is quite unstable; somewhat surprisingly, \autoref{bivartable} shows that our method works well even with $N=25,000$ and $p=5$ meaning that the new method has an advantage over the more restrictive, but correctly specified, model in this case due to computational difficulties of maximum likelihood estimation.  
	\begin{table}[ht]
		\centering
		\begin{tabular}{llllll}
			\toprule
			$\gamma$&$\rho$&	ACRR true&ACRR est& ICRR cor&ICRR rmse \\ 
			\midrule
			1.00&0.25&  2.90 & 2.94 & 0.88 & 1.12 \\ 
			1.75&0.25&  5.35 & 5.08 & 0.88 & 3.74 \\ 
			2.50&0.25&  8.34 & 7.37 & 0.89 & 11.03 \\ 
			1.00&0.40 & 2.90 & 2.82 & 0.86 & 1.06 \\ 
			1.75&0.40&  5.35 & 4.99 & 0.90 & 3.57 \\ 
			2.50&0.40&  8.34 & 6.74 & 0.89 & 12.83 \\ 
			1.00&0.60&   2.90 & 2.77 & 0.83 & 1.18 \\ 
			1.75&0.60&  5.35 & 4.68 & 0.86 & 4.45 \\ 
			2.50&0.60&  8.34 & 6.75 & 0.85 & 13.53 \\ 
			1.00&0.80&  2.90 & 2.23 & 0.61 & 1.82 \\ 
			1.75&0.80& 5.35 & 3.35 & 0.67 & 6.97 \\ 
			2.50&0.80 &8.34 & 4.53 & 0.67 & 18.99 \\
			\bottomrule
		\end{tabular}
		
		\caption{We simulate from the bivariate probit with 25,000 observations and deploy our methodology.  ACRR = average causal risk ratio. ICRR = individual causal risk ratio, cor refers to the correlation between predicted and true for the individual causal risk ratios, and the rmse is the root mean square error.}
		\label{bivartable}
	\end{table}

	\subsection{Sensitivity to $f$}\label{nonlindgp}
	
	We do much better with our methodology when the data were generated from a non-linear data generating process, as described below:
	\begin{align}
		\begin{split}
			b_0(\bm{x})&=\bm{x}_5+\bm{x}_1 \sin(2\bm{x}_6)-1.75\\
			b_1(\bm{x})&=b_0(\bm{x})+1.5\\
			g(\bm{x})&=0.5 b_0(\bm{x})+\bm{x}_2+0.25\\
			U&\sim \mbox{N}(\mu,\sigma^2)\\
			G&\sim \text{Bin}\qty(\Phi(g(\bm{x})+u)) \\
			B\mid G=1&\sim\text{Bin}\qty(\Phi(b_1(\bm{x})+u))\\
			B\mid G=0&\sim \text{Bin}\qty(\Phi(b_0(\bm{x})+u))
		\end{split}
		\label{newnonlinmodel}
	\end{align}
	where we draw $u$ and $b_i(\cdot)$, $G$ conditional on those values, and subsequently  the values of $B$ are drawn conditional on our values of $G$. The $\bm{x}_i$ are drawn uniform($-$1,1), with some $\bm{x}_i$ passed as covariates in our monotone BART fitting stage that do not appear in the DGP; these extraneous variables serve as ``noise'' to complicate the problem and make it more realistic.  \autoref{nonlintable} demonstrates how in this setting our model performs much better than the bivariate probit.
	Additionally, we misspecify $f(u)$ to see if we can still return true individual treatment effects, and, if we fail, what type of distributions cause problems.  In \autoref{nonlintable}, we misspecify with Laplacian distributions, as the fatter tail weight could be problematic, and the table confirms this does appear to be an issue.  Additionally, we compare our methodology with the bivariate probit model, fit with regression spline smoothing and without.  Our methodology does comparatively much better in this setting, as the DGP is highly non-linear. 
	
	In \autoref{nonlintablesharkfin}, we generate $f(u)$ according to the shark fin but with $\sigma$ varied to attain certain variances.  The choice of $q$ affects the skewness of the distribution.  The shark fin provides us insight into whether or not skewness or large variances affect our models estimates; as the previous table showed mean offsets do not seem to impact our estimates too badly.   In \autoref{nonlintablesharkfin_2}, we see getting $q$ wrong (skewness) seems less impactful, meanwhile downwardly estimating variance seems to bias the estimates of the average causal risk ratio (ACRR) up, while guessing variance too high downwardly biases the average causal risk ratio.  \autoref{RRT_RRC_compare} investigates more drastically misspecifying $q$ or $\sigma$.
	
	\begin{table}[h]
		\scalebox{0.9}{
			\begin{tabular}{lP{1.2cm}P{1.2cm}P{.7cm}lP{1.cm}P{.8cm}P{.8cm}P{.8cm}P{.8cm}P{.8cm}}
				\toprule
				$f(u)$& \textbf{true} ACRR&  true est.   ACRR&  RMSE  & Wrong $f(u)$  & Wrong est. &  Wrong RMSE&LBP est.& LBP RMSE&SBP est.&SBP RMSE \\ 
				\midrule
				$\N(0,1)$ &  4.43 & 4.71  &1.66 &Lap(0,$1.2$ )&2.02&3.09&4.19&1.98&4.46&2.00\\
				$\N(0,1.5)$  & 2.80 & 2.81 &0.70 &Lap(0, $1.75)$&1.68&1.36&3.22&0.85&3.38&0.94\\
				$\N(0, 2)$  & 2.14& 2.11 &  0.36& Lap(0, $2.5$)&1.42&0.83&0.44&1.81&0.37&0.73 \\
				$\N(0,2.5 )$  & 1.81 & 1.80 & 0.25&Lap(0, $2$)&2.04&0.37&1.81 &0.46&0.37&1.46\\ 
				$\N(-1, 1)$ & 8.18 & 9.38 & 5.07&Lap($-1$, $1.3$)&1.53&7.83&2.96&6.51&2.83&6.61\\ 
				$\N(1, 2)$  & 1.74 & 1.45 & 0.34&Lap(1, $2.4$)&1.20&0.57&1.89&0.30&0.62&1.15\\ 
				$\N(-2, 2)$  & 3.43 & 5.88 & 4.32&Lap($-2, 2.3$)&1.02&2.49&2.77&0.92&3.03&0.76\\
				$\N(2,1)$  & 1.68 & 1.62&  0.23&Lap($2, 1.3$)&1.18&1.39&0.61&0.45&1.75&1.42\\ \bottomrule
			\end{tabular}
		}
		\caption{Different $f(u)$ as described in the DGP of \autoref{newnonlinmodel}.  $N=25,000$.  Wrong $f(u)$ indicates the distribution of $U$ we used to solve the system of equations in \autoref{maineq} (i.e., how we misspecified).  True indicates true average causal risk ratio (ACRR), and correct est. indicates our estimate of the ACRR when \emph{correctly} specifying $f(u)$. We use standard deviation instead of variance for our spread parameter. Lap refers to the Laplacian distribution. LBP refers to bivariate probit regression without smoothing, and SBP refers to bivariate probit regression with smoothing covariates (i.e., where the smooth term for each covariate is made of basis functions).}
		\label{nonlintable}
	\end{table}
	
	\begin{table}[h]
		\centering
		\scalebox{.9}{
			\begin{tabular}
				{P{2.2cm}P{.6cm}P{1.6cm}P{.8cm}lP{.9cm}P{1.9cm}}
				\toprule
				
				$f(u)$
				sharkfin with parameters $q$, $s$& \textbf{true} ACRR&  true est. ACRR& true  RMSE  & wrong $q$   & wrong $q$ est. &  wrong $q$  RMSE  \\ 
				\midrule
				(0.25, 0.82; 3) &  1.79 & 1.81 &0.21 &(0.40,1.37;3.00)&1.80&0.23\\
				(0.40, 1.37; 3)  & 2.07 & 2.08 &0.31 &(0.70,2.34;3.00)&2.55&0.94 \\
				(0.60, 1.06; 3)  & 3.10 & 2.97 &  0.80& (0.30,1.00;3.00)&1.97&1.43 \\
				(0.75, 2.46; 3)  & 5.86 & 5.37 & 2.59&(0.92,2.77;3.00)&8.41&5.13\\ 
				(0.25, 0.34; 0.5) & 4.11 & 4.27 & 1.54&(0.10,0.12;0.50)&4.13&1.44\\ 
				(0.40, 0.56; 0.5)  & 5.28 & 5.95 & 2.71&(0.20,0.26;0.50)&5.25&2.01\\ 
				(0.60, 0.84; 0.5) & 8.63 & 8.80 &5.31&(0.80,1.05;0.50)&10.9&7.60\\ 
				(0.75, 1.00; 0.5)  & 13.4 & 12.3 & 8.25&(0.45,1.63;0.50)&7.74&10.6\\ \bottomrule
			\end{tabular}
		}
		\caption[Misspecifying distribution: check skewness]{Different $f(u)$ as described in DGP of \autoref{newnonlinmodel}, all of the ``sharkfin'' family.  ACRR = average causal risk ratio. $N=25,000$. Wrong q indicates that we purposely misspecified q when solving our system of equations, whereas the true est. columns indicate where we correctly specified $f(u)$ (both the $q$ and $s$ parameters) when solving our system. ; indicates the variance, whereas the first two entries in shark are the $q$ and $s$ parameters. Here we vary the skewness while keeping variance constant.}
		\label{nonlintablesharkfin}
	\end{table}
	\begin{table}[h]
		\centering
		\scalebox{.9}{
			\begin{tabular}
				{P{2.2cm}P{1.cm}P{1.6cm}P{.8cm}lP{.9cm}P{.9cm}}
				\toprule$f(u)$
				sharkfin with parameters $q$, $s$& \textbf{true} ACRR& true  est. ACRR & true  RMSE   & wrong $\sigma^2$  & wrong $\sigma^2$ est. &  wrong $\sigma^2$  RMSE\\ 
				\midrule
				(0.25, 0.82; 3) &  1.79 & 1.76  &0.38 &(0.25,0.47;1.0)&3.55&2.12\\
				(0.40, 1.37; 3)  & 2.07 & 2.09 &0.61 &(0.40,1.12;2.0)&2.76&0.90 \\
				(0.60, 1.06; 3)  & 3.10 & 3.27 &  1.54& (0.60,0.92;0.6)&11.3&10.4 \\
				(0.75, 2.46; 3)  & 5.86 & 7.40 & 7.94&(0.75,1.74;1.5)&9.79&6.20\\ 
				(0.25, 0.34; 0.5) & 4.11 & 5.34 & 6.25&(0.25,0.67;2.0)&1.56&3.16\\ 
				(0.40, 0.56; 0.5)  & 5.28 & 8.91 & 10.7&(0.40,1.12;2.0)&1.83&4.28\\ 
				(0.60, 0.84; 0.5) & 8.63 & 9.65 & 22.5&(0.60,2.38;4.0)&1.40&9.29\\ 
				(0.75, 1.00; 0.5)  & 13.4 & 16.6 & 57.2&(0.75,3.18;5.0)&1.90&15.5\\ \bottomrule
			\end{tabular}
		}
		\caption[Misspecifying distribution: check variance change]{Different $f(u)$ as described in DGP of \autoref{newnonlinmodel}, all of the ``sharkfin'' family.  $N=25,000$. Wrong $\sigma^2$ indicates that we purposely misspecified our variance (by varying the $\sigma$ parameter) when solving our system of equations, whereas the true est. columns indicate where we correctly specified $f(u)$ (both the $q$ and $s$ parameters) when solving our system. ; indicates the variance, whereas the first two entries in shark are the $q$ and $s$ parameters. Here we vary the variance keeping skewness constant.  }
		\label{nonlintablesharkfin_2}
	\end{table}
	
	\begin{table}[ht]
		\centering
		\scalebox{0.9}{
			\begin{tabular}{lP{1.cm}P{1.cm}P{1.cm}P{1.cm}lP{1.6cm}P{1.6cm}}
				\toprule
				
				True $f(u)$ & true ACRRT   &ACRRT est.&ACRRC true &ACRRC est.& Wrong q $f(u)$&ACRRT est. wrong& ACRRC est. wrong\\ 
				\midrule 
				shark(0.1, 0.30; 3) & 1.60 &1.62&1.67&1.69 &shark(0.9, 2.74;3)&1.36&1.60 \\ 
				shark(0.1, 0.12; 0.5) & 3.18&3.16&3.79& 3.75&shark(0.9, 1.12; 0.5)&3.82&5.19 \\
				shark(0.1, 0.18; 1)&2.32&2.39&2.58&2.69&shark(0.9, 1.58; 1)&2.75&3.72\\
				shark(0.1, 0.18; 1)&2.32&2.39&2.58&2.69&shark(0.5, 1; 1)&2.43&2.97\\
				shark(0.5, 1; 1)&4.00&4.18&4.66&5.18&shark(0.1, 0.18; 1)&3.16&3.47\\
				shark(0.5, 1; 1)&4.00&4.18&4.66&5.18&shark(0.9, 1.58; 1)&6.78&9.32\\
				shark(0.9, 1.58; 1)&13.1&12.0&19.2&17.6&shark(0.1, 0.18; 1)&2.75&2.56\\
				shark(0.9, 1.58; 1)&13.1&12.0&19.2&17.6&shark(0.5, 1; 1)&4.96&5.66\\
				\bottomrule
			\end{tabular}
		}
		\caption[Misspecifying distribution: check skewness for bigger q, ACRRT and ACRRC]{Comparing estimates of average causal risk ratio on treated (ACRRT) and average causal risk ratio on controls (ACRRC) when we more aggressively misspecify the q parameter, which controls the skewness. }
		\label{RRT_RRC_compare}
	\end{table}

	\subsection{Relationship with E-values: Simulations} \label{E-val_sec}

\begin{figure}[t]
		\centering
		
		\includegraphics[height=8.25cm]{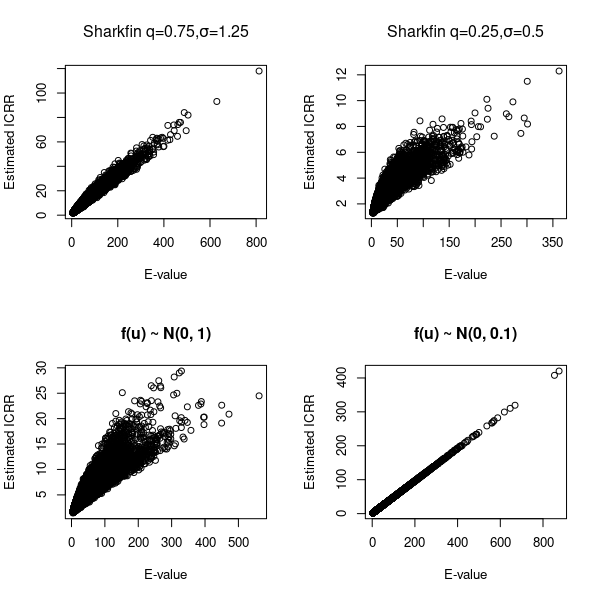}
		
		\caption[E-val vs ratio]{  Comparison of our individual causal risk ratio estimates vs. individual E-value estimates for 25,000 simulations drawn from the same dgp as specified in \autoref{newnonlinmodel}.  Shown are different distributions of $f(u)$, with the bottom right ``low u'' setting recapitulating the E-value. }
		\label{E-val-ratio}
		
	\end{figure}

	Here, we replicate the analysis presented in \autoref{E-val-ratio_audit} with simulated data.  Rather than using all the posterior draws given by the BART model in the simulated data setting, we instead take the mean of the posterior BART probability estimates in the fitting stage and then solve for the causal parameters $b_0(\vdot), b_1(\vdot), g(\vdot)$ once for each observation.  We impose the constraint that $b_1(\bm{x})\geq b_0(\bm{x})$ when solving for the causal parameters. We do this for different distributions of $f$ with the data generated according to  \autoref{newnonlinmodel}.   In \autoref{E-val-ratio}, we compare our estimate of the inducement effect vs. the E-value, for different distributions of $U$. For choices of $f$ that concentrate near zero, the estimated individual causal risk ratios effectively recapitulate the E-values, while for choices of $f$ that entail higher probability of relevant unobserved private information the estimates differ from the E-value in ways that depend on the specific shape of $f$. 
	
	\subsection{Value of monotonicity}\label{mono_section}
\begin{figure}[t]
		\centering
		\begin{subfigure}{.4\textwidth}
			\includegraphics[width=5.cm]{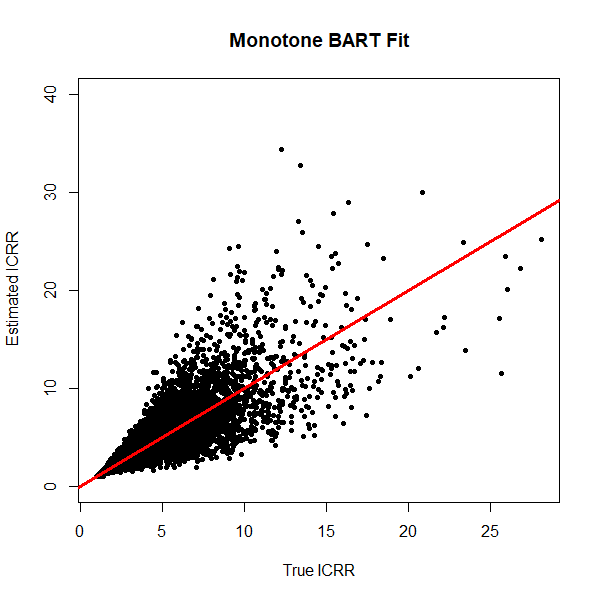}
		\end{subfigure}%
		\begin{subfigure}{.4\textwidth}
			\includegraphics[width=5.cm]{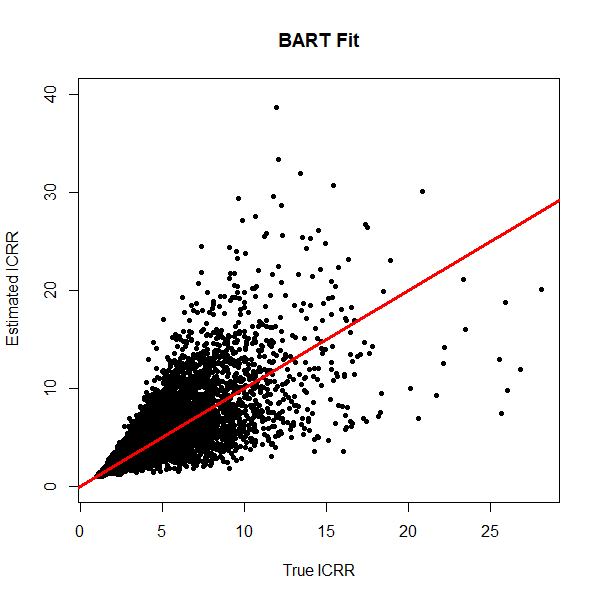}
		\end{subfigure}
		\caption[Comparing Bart with and without monotonicity constraint]{Plots of expected individual causal risk ratios vs. our estimates; i.e., a plot comparing the ratio of potential outcomes from the model described in \autoref{latentutility} ($\Phi(\alpha_0+\alpha_1\bm{x}_i+\gamma)/\Phi(\alpha_0+\alpha_1\bm{x}_i)$) versus our estimate within the integral of \autoref{treateq}.  In the DGP, $\rho=0.25,\gamma=1$.  The monotone BART correlation between the truth and estimate is 0.88 and for BART it is 0.83. We simulate 25,000 samples.}
		\label{monovsnorm}
	\end{figure}

	\autoref{monovsnorm} compares estimates of the individual causal risk ratios under a BART model with versus without monotonicity. As expected, because the monotonicity constraint is satisfied in this data generating process, the model that imposes that restriction exhibits greater accuracy.
	
	\section{Discussion}

	Compared to the popular bivariate probit model, the machine learning sensitivity analysis introduced here is more flexible and hence, more credible in empirical analyses. This increased flexibility comes at the price of identification, but this should not be a barrier to empirical investigation: a thorough sensitivity analysis can still yield evidence and insight, especially when coupled with posterior subgroup analysis.
	
	Specifically, we conclude that at least some firms appear to experience induced bankruptcies; the degree of private information would have to be extreme to rule this out entirely. Moreover, it appears that induced bankruptcies are more likely to occur for firms that have high levels of leverage and that have an S\&P credit rating. These results are reassuring given that adverse going concern opinions can mechanically lead to higher borrowing costs and credit rating downgrades. The fact that these moderating variables were uncovered by the model without explicit instruction lends credence to the inducement hypothesis.

	Data analyses which mirror the ``self-fulfilling prophecy'' of the bankruptcy inducement problem have the potential to benefit from the modular machine learning sensitivity analysis developed here. For example, the question of whether Catholic high schools lead to higher college enrollment \citep{Evans-Schwab-1995} would be of particular interest, as that analysis employed the bivariate probit with endogenous regressor approach that we have generalized.

	Another area of future research would be to allow for the distributions of $U$ to be dependent across firms. For an individual firm, the interpretation of $U$ is indeed dictated by the choice of the unidentifiable density function $f$. However, we are making a substantive assumption in this paper that the distribution of private information between firms is uncorrelated; this assumption is what allows us to solve the equations one by one for each firm.
	In one sense, this assumption is clearly unreasonable, as unobservable shocks to industries could affect all firms in that industry. However, such modeling would be entirely assumption-driven (since $U$ is unobserved) and would not, we surmise, affect the marginal point estimates much. Permitting strong dependencies would, we suppose, affect the resulting uncertainty intervals, but at the cost of an infeasible computational burden.

	In \autoref{section:audit_endo}, we study the usefulness of a rich economic model that explicitly accounts for auditor behavior economically.  We suggest estimating such a model as a future area of research.  Discussion of this model is left in the appendix as it conflicts with our general approach, which is agnostic about details of the data generating process. Moreover, it is unclear whether auditors are actively trying to avoid inducement effects. In this regard, what our approach is measuring is the observed inducement effect and not the counterfactual inducement effect if auditors had accounted for it when considering the issuance of a going concern opinion.

	\color{black}

	\section*{Acknowledgements}
	The authors would like to acknowledge support from NSF grant \#1502640.  The authors also thank ASU Research Computing facilities for providing computing resources.  
	Thanks are also in order to Samantha Brozak, Andrew Herren, and Chelsea Krantsevich for helpful feedback.  


	\bibliography{bankruptcy_sensitivity_final_version}
	\clearpage
	\appendix
	\section{MCMC Diagnostics}\label{appendix_mcmc}

	We run the monotone BART model with mostly the default specifications of \cite{bart} but with 2,000 burn-in draws, 2,000 posterior draws, and 100 trees (the specifications we use throughout in the simulated data and empirical analysis).  The BART model in the convergence comparison uses 1,000 cut-points generated uniformly, as was consistent throughout the paper. Convergence diagnostics are presented in \autoref{mcmc_diagnostics}.
	\begin{table}[h]
		\centering
		\begin{tabular}{lP{1.5cm}P{1cm}}
			
			\toprule
			Name&Geweke Diagnostic& $n_{\text{eff}}$\\ \midrule
			Bart $\Pr(B\mid G=0, \bm{x})$&0.27&40\\
			Bart $\Pr(B\mid G=1, \bm{x})$&0.40&261\\
			Monotone Bart $\Pr(B\mid G=0, \bm{x})$&0.24&46\\
			Monotone Bart $\Pr(B\mid G=1, \bm{x})$&0.30&187\\
			
			\bottomrule      
		\end{tabular}
		\caption{The $\Pr(B\mid G=0, \bm{x})$ rows refer to training on (only) firms that did not receive a going concern and predict on the entire dataset.  Similarly, $\Pr(B\mid G=1, \bm{x})$ refers to only training on firms that received a going concern. The effective sample size is estimated such that  $\text{Var}_{\text{MCMC}}(\overline{\vdot})=\frac{\text{Var}(\vdot )}{n_{\text{eff}}}$.  The Geweke diagnostics refers to the convergence diagnostics of \cite{geweke1992evaluating}, which returns a Z-score (which we convert to the probability scale) for test of equality of means between the first and last parts of the MCMC chain. We present the mean of these diagnostics across 1,000 of the firms in the dataset, checking for the convergence in the posterior predictions.  }
		\label{mcmc_diagnostics}
	\end{table}

\begin{figure}[!httb]
	\begin{subfigure}{.5\textwidth}
		\includegraphics[width=6.2cm]{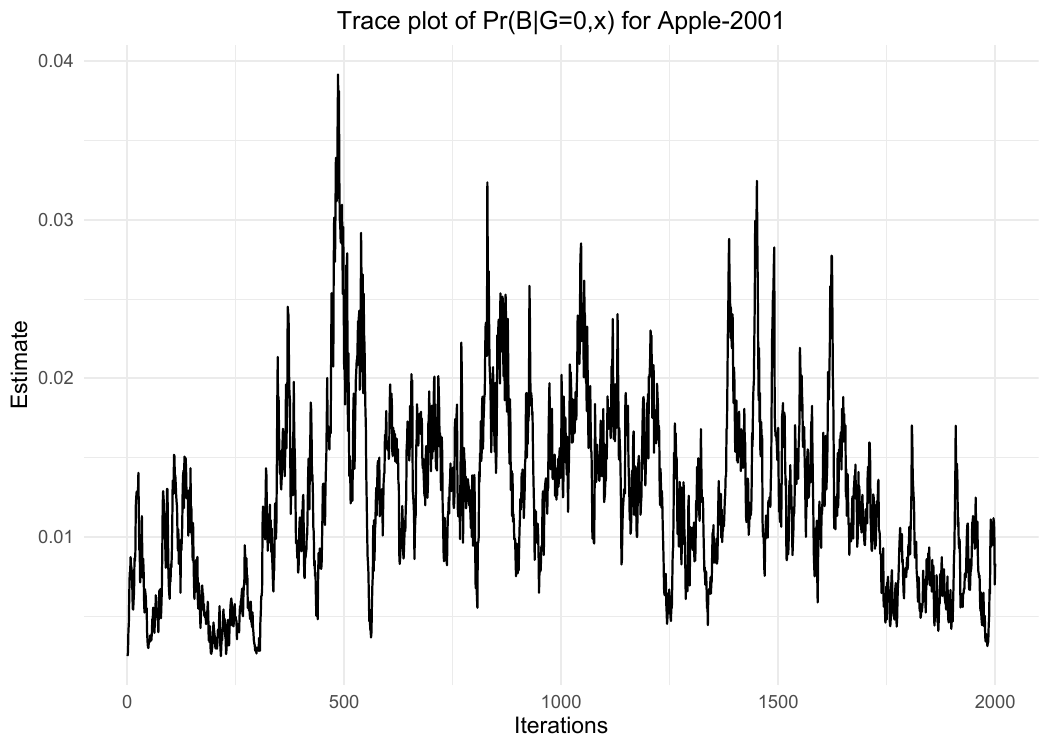}
	\end{subfigure}%
\begin{subfigure}{.5\textwidth}
	\includegraphics[width=6.2cm]{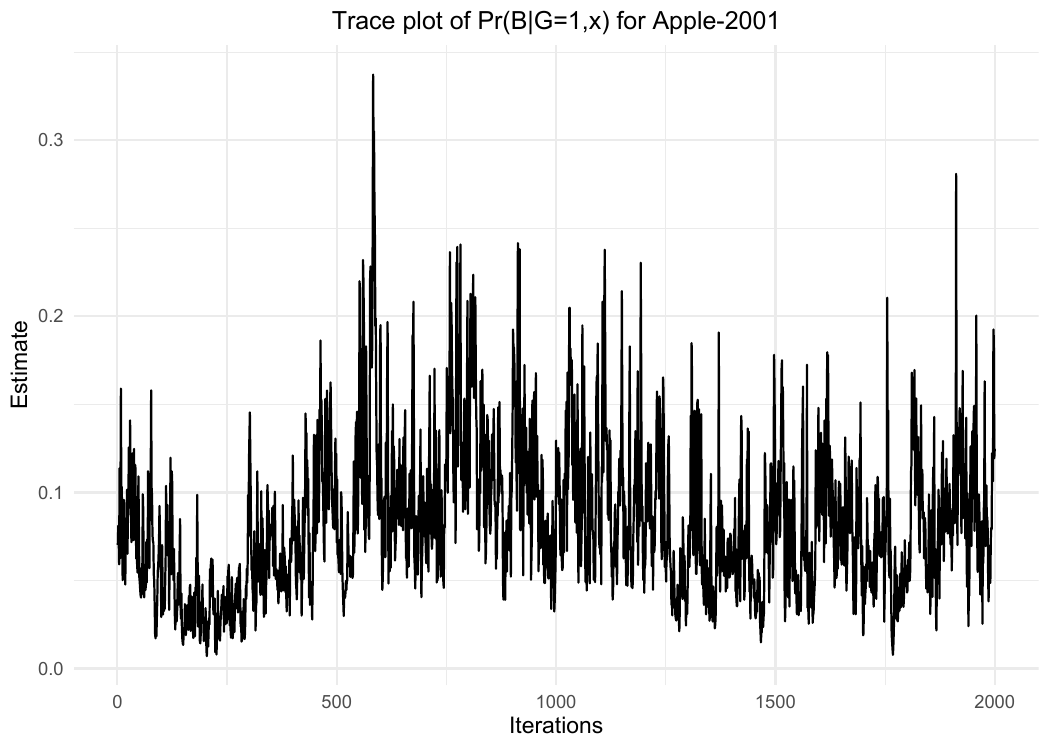}
\end{subfigure}
	\begin{subfigure}{.5\textwidth}
	\includegraphics[width=6.2cm]{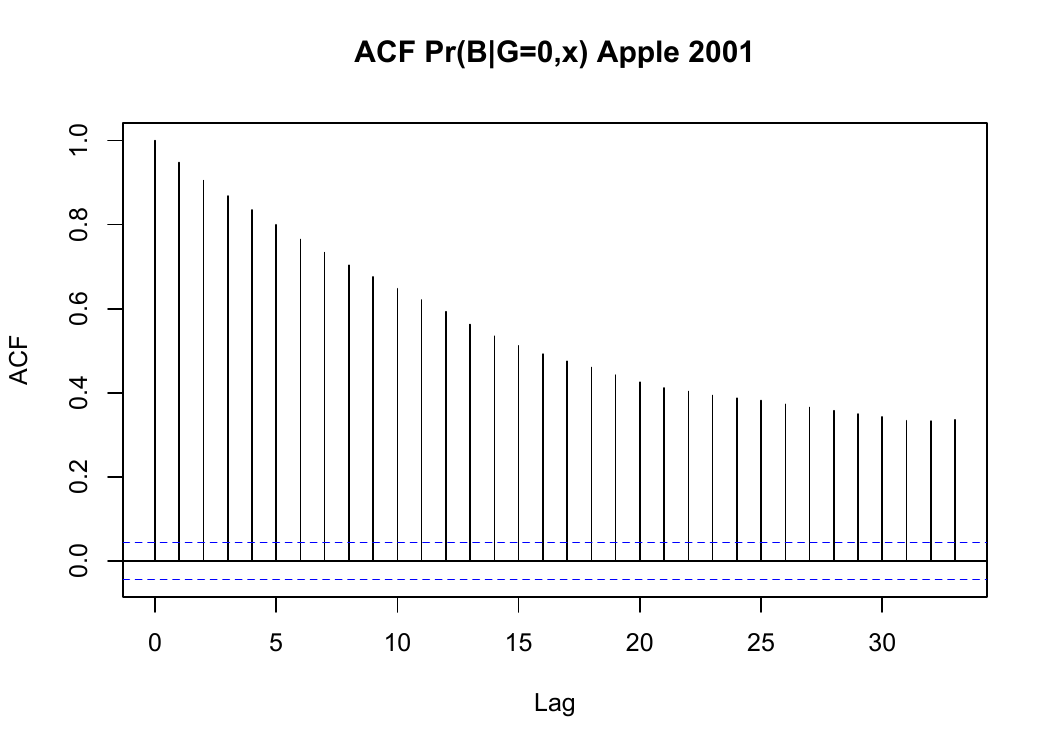}
\end{subfigure}%
\begin{subfigure}{.5\textwidth}
	\includegraphics[width=6.2cm]{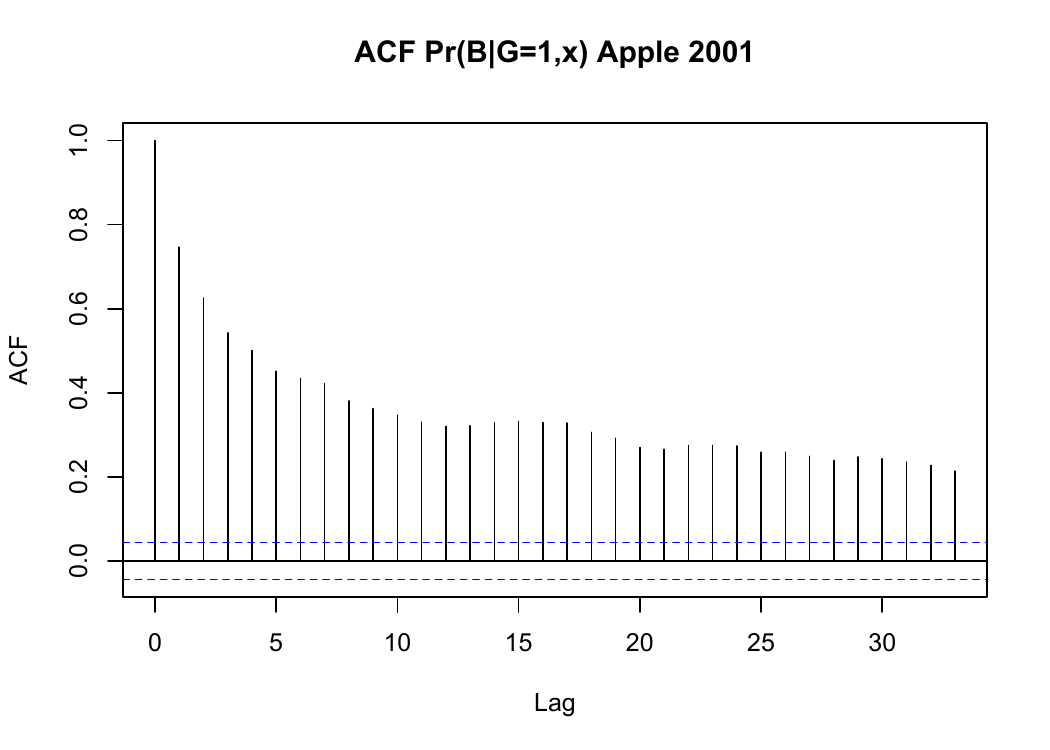}
\end{subfigure}

\caption{Top row are trace plots for the posterior (after 2,000 burn-in samples) of the Apple-2001 firm year observation.  On the left is a trace plot for the monotone BART estimate of $\Pr(B\mid G=0, \bm{x})$, the right is $\Pr(B\mid G=1, \bm{x})$.  Bottom row are plots of auto-correlation for the MCMC chain.}

\end{figure}

 \section{Auditors Modeling Endogeneity}\label{section:audit_endo}

\begin{enumerate}
	
	\item \textit{Why auditors do not avoid inducement effects.}

	Inducement effects occur when lenders (e.g., banks), observing a going concern opinion, force borrowers into bankruptcy. Auditors cannot prevent such inducement effects for at least two reasons. 
	
	First, auditors have an asymmetric loss function. An auditor who fails to issue an adverse going concern opinion for a client that subsequently goes bankrupt (type 2 error) will face significant litigation risk. In contrast, issuing a going concern opinion for a client that does not go bankrupt (type 1 error) has lower litigation concerns, because the auditors' job is to evaluate whether there is substantial doubt about the entity's ability to continue as a going concern for a reasonable period, not to predict bankruptcy. Due to this asymmetric loss function, auditors have strong incentives to issue a going concern opinion prior to bankruptcy to reduce their exposure to litigation risk and lower settlement amounts (see \cite{kaplan2013going}), implying that their incentive to prevent bankruptcy is not strong. 
	
	Second, whether a firm goes bankrupt depends on the bargaining outcome with its lenders. It is difficult for auditors to directly interfere with this bargaining process and prevent bankruptcy from happening.

	\item \textit{A structural model of going concern opinion.}

	The auditor chooses to issue a going concern opinion $g\in\{0,1\}$ by maximizing its expected utility (we drop the conditioning variable $X_i$ to economize notation):
	\begin{equation}
		\max_{g\in\{0,1\}} \E \bigg[ g*(1-B) u_{1} + gB u_{2} + (1-g) B u_{3} + (1-g) (1-B) u_{4}\bigg], \label{eq:Obj}
	\end{equation}
	where $B$ is an indicator for bankruptcy and $u_{i\in\{1,...,4\}}$ are the utilities for the auditor that depend on $X_i$, $g$, and $B$. 
	
	In what follows, we normalize the utilities corresponding to a correct going concern opinion to zero, that is, $u_{2}=u_{4}=0$. A going concern opinion is issued if and only if
	\begin{equation}
		G =1 \Leftrightarrow  \E \bigg[ (1-B) u_{1} |G =1 \bigg] \geq \E \bigg[ B u_{3} | G=0  \bigg] . \label{eq:GCModel}
	\end{equation}
	
	The model is completed by noting that bankruptcy occurs when
	\begin{equation}
		\epsilon_b \leq \gamma_0 + \gamma_1 G.
	\end{equation}
	The probability of a bankruptcy equals $\Phi(\gamma_{0} + \gamma_{1} G)$, where $\Phi(\cdot)$ is the CDF of a standard normal distribution. Substituting it into \autoref{eq:GCModel}, we know that a going concern opinion is issued if and only if
	\begin{equation}
		(1- \Phi(\gamma_0 + \gamma_1)) u_{1}  -  \Phi(\gamma_0)u_{3} \geq 0. \label{eq:GCModel1}
	\end{equation}
	
	The auditor uses a threshold strategy to issue a going concern opinion. Our discussion below assumes that $u_1<0$ and $u_3<0$, which is reasonable in that they capture the utilities to the auditor when issuing an ``incorrect'' going concern opinion that differs from the bankruptcy outcome. As shown in \autoref{eq:GCModel1}, the auditor is more likely to issue a going concern opinion when the cost of type 1 error, $|u_{1}|$, is small, relative to the cost of type 2 error, $|u_{3}|$. 
	
	The discussion above assumes that the auditor internalizes the inducement effect of a going concern opinion. On the one hand, the auditors' job is to evaluate whether there is substantial doubt about the entity's ability to continue as a going concern for a reasonable period. If a going concern opinion can induce bankruptcy, this effect should be accounted for. On the other hand, internalizing the inducement effect implies a greater likelihood of a going concern opinion. This may cause client objection, which may discourage the auditor from internalizing the inducement effect. This is plausible as going concern opinions are issued after evaluating business plans proposed by the client. In this case, a going concern opinion is issued if and only if 
	\begin{equation}
		(1- \Phi(\gamma_0)) u_{1}  -  \Phi(\gamma_0)u_{3}  \geq 0. \label{eq:GCModel2}
	\end{equation}
	
	Our structural model captures the utility difference in \autoref{eq:GCModel1} or \autoref{eq:GCModel2} using observable client firm characteristics. 
	The benefit of this structural model is that researchers can separately estimate the sources of inducement effects driven by auditors' concerns of type 1 and type 2 errors. Estimating this model requires data that can be used to separate the costs of type 1 and type 2 errors for auditors (e.g., auditor litigation data).

\end{enumerate}	

	\section{Estimating the risk difference as estimand of interest}\label{ARD_focus}
Rather than looking at the ratio of potential outcomes, it is often the case we want to investigate the difference in the expected value of each, i.e. we can look at \emph{risk differences}:
\[\text{Risk Difference}\rightarrow \Delta \equiv \E(B^1)-E(B^0)\]
In our framework, following similar reasoning as in \autoref{section_indirect_inference},  risk differences can be defined as 
\[\Delta(\mathbf{x}_i)=\int_{\mathbb{R}} \Phi\qty(b_1(\mathbf{x})+u)f(u)\dd u - \int_{\mathbb{R}} \Phi\qty(b_0(\mathbf{x})+u) f(u)\dd u\]
The sample average risk difference (ARD) is therefore $\frac{1}{n}\sum_{i=1}^{n}\Delta(\mathbf{x}_i)$. In the case of the audit data, the average risk difference refers to percentage point difference in going bankrupt after receiving a going concern opinion. We esimate the risk difference using our methodology as well as the bivariate probit with endogenous regressor model, (described in equation(4) in the main file), to the audit data.  Specifically, we used the same covariates as we used when fitting monotone bart, used the bankruptcy indicator as the binary outcome, and whether or not a going concern was issued as the ``treatment'' indicator.  Using our methodology, results for estimating risk differences on the audit data are presented in  \autoref{resultssummary}.

Analogs of \autoref{nonlintable},
\autoref{nonlintablesharkfin}, \autoref{nonlintablesharkfin_2}, and \autoref{RRT_RRC_compare} with the causal risk difference as the estimand of interest  are presented in \autoref{nonlintable_app},
\autoref{nonlintablesharkfin_app}, \autoref{nonlintablesharkfin_2_app}, and \autoref{ATT_ATC_compare}.

\begin{table}[!httb]
	\centering
	
	\begin{tabular}{llllc}
		
		\toprule
		Distribution of $f(u)$  & 
		&ARD post (\%)& mean $B_1$ (\%) &95\% Credible interval for ARD (\%)  \\ \midrule
		$\N(0,\sigma=0.1)$ &
		&9.97&11.5&$\qty(7.08, 12.9)$\\ 
		$\N(0,\sigma=0.5)$ &
		&4.12 &5.96 &$\qty(2.74, 5.60)$\\
		$\N(0,\sigma=1)$&
		&0.70&2.90&$\qty(0.40, 1.01)$\\
		Shark $q=0.25$, $s=0.5$; $\sigma=1.05$&
		&0.28&2.60&$\qty(0.14, 0.45)$\\
		Shark $q=0.75$, $s=1.25$; $\sigma=0.89$&
		&2.84&4.87&$\qty(1.81, 3.97)$\\
		
		Symmetric Mixture ($\sigma=0.64$)&
		&2.30&4.52&$\qty(2.23, 2.92)$\\
		
		Asymmetric Mixture ($\sigma=0.49$)&
		&2.47&4.62&$\qty(2.43, 3.14)$\\

		\bottomrule
		
	\end{tabular}
	\caption{The reduced form probabilities (\autoref{long} in the main file) were estimated using BART with a monotonicity constraint on the going concern variable.  We further require $b_1(\bm{x})>b_0(\bm{x})$ in the projection step.  Posterior summaries based on 500 Monte Carlo samples.  $\sigma$ refers to the implied standard deviations of the different distributions.  Values listed as the percentage increase in probability of bankruptcy.}
	\label{resultssummary}
	
\end{table}

\begin{table}[h]
	
	\begin{tabular}{lllP{.8cm}lP{1.cm}P{.8cm}P{.8cm}P{.8cm}P{.8cm}P{.8cm}}
		\toprule
		$f(u)$& \textbf{true} ARD& Correct est.& Correct RMSE  & Wrong $f(u)$  & Wrong est. &  Wrong RMSE&LBP est.& LBP RMSE&SBP est.&SBP RMSE \\ 
		\midrule
		$N(0,1)$ &  0.30 & 0.31  &0.05 &Lap(0,$1.2^2$ )&0.17&0.15&0.14&0.16&0.16&0.16\\
		$N(0,1.5^2)$  & 0.26 & 0.26 &0.05 &Lap(0, $1.75^2)$&0.16&0.12&0.13&0.13&0.06&0.19 \\
		$N(0, 2^2)$  & 0.23 & 0.23 &  0.04& Lap(0, $2.5^2$)&0.12&0.12&0.23&0.06&0.25&0.07 \\
		$N(0,2.5^2 )$  & 0.20 & 0.20 & 0.04&Lap(0, $2^2$)&0.25&0.&0.07 &0.07&0.23&0.06\\ 
		$N(-1, 1)$ & 0.17 & 0.18 & 0.04&Lap(-1, $1.3^2$)&0.05&0.14&0.10&0.16&0.11&0.16\\ 
		$N(1, 2^2)$  & 0.25 & 0.22 & 0.06&Lap(1, $2.4^2$)&0.11&0.16&0.17&0.06&0.16&0.08\\ 
		$N(-2, 2^2)$  & 0.11 & 0.03 & 0.09&Lap($-2, 2.3^2$)&0.00&0.12&0.07&0.08&0.07&0.09\\ 
		$N(2,1)$  & 0.30 & 0.29 &  0.05&Lap($2, 1.3^2$)&0.10&0.22&0.28&0.13&0.24&0.13\\ \bottomrule
	\end{tabular}
	\caption[Mis-specifying distribution: check mean and variance shifts, tail weights]{Different $f(u)$ as described in \autoref{newnonlinmodel} of the main file.  $N=25000$. Wrong $f(u)$ indicates the distribution of $U$ we used to solve the system of equations in \autoref{long} , i.e. how we mis-specified.  True indicates true ARD, and correct est indicates our estimate of the ARD when \emph{correctly} specifying $f(u)$. Lap refers to the Laplacian distribution. Smooth refers to bivariate probit regression with smoothing covariates.}
	\label{nonlintable_app}
\end{table}

\begin{table}[h]
	\centering
	\begin{tabular}
		{P{2.6cm}P{.6cm}P{1.2cm}P{.8cm}lP{1.5cm}P{1.5cm}}
		\toprule
		
		$f(u)$
		sharkfin with parameters $q$, $s$& \textbf{true} ARD& true  est. ARD& true  RMSE  & wrong $q$   & wrong $q$  ARD est. &  wrong $q$  RMSE  \\ 
		\midrule
		(0.25, 0.82; 3) &  0.28 & 0.28  &0.04 &(0.40,1.37;3.00)&0.22&0.05\\
		(0.40, 1.37; 3)  & 0.26 & 0.27 &0.05 &(0.70,2.34;3.00)&0.29&0.05\\
		(0.60, 1.06; 3)  & 0.23 & 0.22 &  0.04& (0.30,1.00;3.00)&0.15&0.07 \\
		(0.75, 2.46; 3)  & 0.18 & 0.17 & 0.04&(0.92,2.77;3.00)&0.21&0.08\\ 
		(0.25, 0.34; 0.5) & 0.38 & 0.39 & 0.05&(0.10,0.12;0.50)&0.39&0.05\\ 
		(0.40, 0.56; 0.5)  & 0.35 & 0.36 & 0.06&(0.20,0.26;0.50)&0.35&0.06\\ 
		(0.60, 0.84; 0.5) & 0.29 & 0.30 & 0.05&(0.80,1.05;0.50)&0.31&0.06\\ 
		(0.75, 1.00; 0.5)  & 0.24 & 0.25 & 0.05&(0.45,1.63;0.50)&0.21&0.06\\ \bottomrule
	\end{tabular}
	\caption[Mis-specifying distribution: check skewness]{Different $f(u)$ as described in \autoref{newnonlinmodel}, all of the ``sharkfin'' family.  $N=25000$. Wrong $q$ indicates that we purposely mis-specified $q$ when solving our system of equations, whereas the true est. AR and true RMSE columns indicate where we correctly specified $f(u)$, both the $q$ and $s$ parameters, when solving our system. ; indicates the variance, whereas the first 2 entries in shark are the $q$ and $s$ parameters. Here we vary the skewness while keeping variance constant.}
	\label{nonlintablesharkfin_app}
\end{table}
\begin{table}[h]
	\centering
	\begin{tabular}
		{P{2.6cm}P{.6cm}P{1.2cm}P{.8cm}lP{1.5cm}P{1.5cm}}
		\toprule$f(u)$
		sharkfin with parameters $q$, $s$& \textbf{true} ARD& true  est. ARD& true  RMSE   & wrong $\sigma^2$  & wrong $\sigma^2$  ARD est. &  wrong $\sigma^2$  RMSE\\ 
		\midrule
		(0.25, 0.82; 3) &  0.28 & 0.28  &0.06 &(0.25,0.47;1.0)&0.50&0.23\\
		(0.40, 1.37; 3)  & 0.26 & 0.27 &0.06 &(0.40,1.12;2.0)&0.35&0.10 \\
		(0.60, 1.06; 3)  & 0.23 & 0.22 &  0.06& (0.60,0.92;0.6)&0.48&0.27 \\
		(0.75, 2.46; 3)  & 0.18 & 0.17 & 0.05&(0.75,1.74;1.5)&0.26&0.09\\ 
		(0.25, 0.34; 0.5) & 0.38 & 0.39 & 0.07&(0.25,0.67;2.0)&0.15&0.24\\ 
		(0.40, 0.56; 0.5)  & 0.35 & 0.36 & 0.07&(0.40,1.12;2.0)&0.13&0.23\\ 
		(0.60, 0.84; 0.5) & 0.29 & 0.30 & 0.07&(0.60,2.38;4.0)&0.04&0.28\\ 
		(0.75, 1.00; 0.5)  & 0.24 & 0.25 & 0.06&(0.75,3.18;5.0)&0.04&0.24\\ \bottomrule
	\end{tabular}
	\caption[Mis-specifying distribution: check variance change]{Different $f(u)$ as described in \autoref{newnonlinmodel}, all of the ``sharkfin'' family.  $N=25000$. Wrong $\sigma^2$ indicates that we purposely mis-specified our variance (by varying the $s$ parameter) when solving our system of equations, whereas the true est. ARD and true RMSE columns indicate where we correctly specified $f(u)$, both the $q$ and $s$ parameters, when solving our system. ; indicates the variance, whereas the first 2 entries in shark are the $q$ and $s$ parameters. Here we vary the variance keeping skewness constant.  }
	\label{nonlintablesharkfin_2_app}
\end{table}

\begin{table}[ht]
	\centering
	\begin{tabular}{lP{1.cm}P{1.cm}P{1.cm}P{1.cm}lP{1.6cm}P{1.6cm}}
		\toprule
		
		True $f(u)$ & true ACRDT  &ACRDT est.&ACRDC true &ACRDC est.& Wrong $q$ $f(u)$&ACRDT est. wrong& ACRDC est. wrong\\ 
		\midrule 
		shark(0.1, 0.30; 3) & 0.28 &0.29&0.29&0.29 &shark(0.9, 2.74;3)&0.14&0.20 \\ 
		shark(0.1, 0.12; 0.5) & 0.40&0.40&0.38& 0.39&shark(0.9, 1.12; 0.5)&0.39&0.40 \\
		shark(0.1, 0.18; 1)&0.37&0.37&0.37&0.36&shark(0.9, 1.58; 1)&0.32&0.37\\
		shark(0.1, 0.18; 1)&0.37&0.37&0.37&0.36&shark(0.5, 1; 1)&0.34&0.36\\
		shark(0.5, 1; 1)&0.32&0.31&0.28&0.29&shark(0.1, 0.18; 1)&0.29&0.24\\
		shark(0.5, 1; 1)&0.32&0.31&0.28&0.29&shark(0.9, 1.58; 1)&0.35&0.34\\
		shark(0.9, 1.58; 1)&0.21&0.20&0.16&0.16&shark(0.1, 0.18; 1)&0.11&0.06\\
		shark(0.9, 1.58; 1)&0.21&0.20&0.16&0.16&shark(0.5, 1; 1)&0.15&0.10\\
		\bottomrule
	\end{tabular}
	\caption[Mis-specifying distribution: check skewness for bigger q, ATC and ATT]{Comparing estimates of average causal risk difference on treated (ACRDT) and average causal risk difference on controls (ACRDC) when we more aggressively misspecify the $q$ parameter, which controls the skewness.}
	\label{ATT_ATC_compare}
\end{table}

\section{Bivariate probit simulation study}\label{bivar_append}
\autoref{bivarvalidate} shows the results when fitting the bivariate probit regression with a maximum likelihood estimate to the simulated bivariate probit data.  Unsurprisingly, this performs well, with the caveat that we require large $\N$ ($\N=100,000$) to get these impressive results.  We simulated the samples from the bivariate probit model of the main document, where we sum 5 uniform(-1,1) $\bm{x}_i$ covariates each with the same $\beta_1$ and $\alpha_1$ coefficients respectively.  We set $\beta_0=0, \beta_1=-0.2, \alpha_0=-0.5, \alpha_1=0.7$ to generate reasonable number of going concerns and bankruptcies.

\autoref{bivartable_treat} shows the results when we simulated from the bivariate probit and fit with our methodology, with $f(u)$ assigned appropriately, only this time we are interested in the treatment effect.  Our method does well here, with $\N=25,000$ and $p=5$.

\begin{table}[t]
	\centering
	\scalebox{.85}{
		\begin{tabular}{lllP{1.3cm}P{1.3cm}P{1.3cm}P{1.3cm}P{1.3cm}llll}
			\toprule
			ACRD true & ACRD est&ICRD cor&ICRD RMSE&ACRR true&ACRR est& ICRR cor&ICRR rmse&  $\gamma$ true&$\gamma$ est.& $\rho$& $\rho$ est. \\ 
			\midrule
			0.23 & 0.24 & 0.97 & 0.02 & 2.24 & 2.07 & 1.00 & 0.24 & 1.00 & 0.77 & 0.25 & 0.37 \\ 
			0.46 & 0.46 & 0.99 & 0.02 & 4.32 & 3.89 & 1.00 & 0.87 & 1.75 & 1.62 & 0.25 & 0.31 \\ 
			0.58 & 0.57 & 1.00 & 0.02 & 6.24 & 5.40 & 0.99 & 2.34 & 2.50 & 2.38 & 0.25 & 0.30 \\ 
			0.26 & 0.26 & 0.99 & 0.01 & 2.42 & 2.27 & 1.00 & 0.22 & 1.00 & 0.85 & 0.40 & 0.47 \\ 
			0.46 & 0.46 & 0.99 & 0.02 & 4.33 & 3.89 & 1.00 & 0.90 & 1.75 & 1.63 & 0.40 & 0.45 \\ 
			0.57 & 0.56 & 0.99 & 0.02 & 6.14 & 5.13 & 1.00 & 2.83 & 2.50 & 2.34 & 0.40 & 0.46 \\ 
			0.28 & 0.28 & 0.99 & 0.01 & 2.57 & 2.46 & 1.00 & 0.17 & 1.00 & 0.92 & 0.60 & 0.63 \\ 
			0.47 & 0.47 & 1.00 & 0.01 & 4.51 & 4.24 & 1.00 & 0.63 & 1.75 & 1.70 & 0.60 & 0.61 \\ 
			0.59 & 0.58 & 1.00 & 0.01 & 6.41 & 5.89 & 0.99 & 1.67 & 2.50 & 2.45 & 0.60 & 0.61 \\ 
			0.31 & 0.31 & 1.00 & 0.00 & 2.79 & 2.80 & 1.00 & 0.02 & 1.00 & 1.02 & 0.80 & 0.80 \\ 
			0.47 & 0.47 & 1.00 & 0.01 & 4.51 & 4.26 & 1.00 & 0.56 & 1.75 & 1.70 & 0.80 & 0.81 \\ 
			0.58 & 0.58 & 1.00 & 0.01 & 6.31 & 5.56 & 0.99 & 2.25 & 2.50 & 2.41 & 0.80 & 0.82 \\ 
			\bottomrule
		\end{tabular}
	}
	\caption[Bivariate probit regression validation]{N=100,000. Fit the simulated bivariate probit with the bivariate probit regression.  Validates the MLE of the bivariate probit regression performs well, as well as the validity of our data generation process, however required a large $N$ to get accurate results. ACRD refers to average causal risk difference, ICRD individual causal risk difference.  ACRR refers to average causal risk ratio, whereas ICRR refers to individual causal risk ratio. }
	\label{bivarvalidate}
\end{table}

\begin{table}[ht]
	\centering
		\begin{tabular}{llllll}
			\toprule
			$\gamma$&$\rho$& ACRD true & ACRD est&ICRD cor&ICRD RMSE \\ 
			\midrule
			1.00&0.25& 0.29 & 0.29 & 0.89 & 0.05  \\ 
			1.75&0.25& 0.47 & 0.47 & 0.96&0.04  \\ 
			2.50&0.25& 0.58 & 0.57 &  0.97&0.05  \\ 
			1.00&0.40 &0.29 & 0.29 &  0.90&0.05  \\ 
			1.75&0.40& 0.47 & 0.46 &  0.94&0.06  \\ 
			2.50&0.40& 0.58 & 0.55 &  0.96&0.09  \\ 
			1.00&0.60&  0.29 & 0.29 &  0.90&0.05  \\ 
			1.75&0.60& 0.47 & 0.46 &  0.95 &0.05 \\ 
			2.50&0.60& 0.58 & 0.57 & 0.98&0.04  \\ 
			1.00&0.80& 0.29 & 0.27 &  0.91&0.05  \\ 
			1.75&0.80& 0.47 & 0.44 &  0.95 &0.05 \\ 
			2.50&0.80 &0.58 & 0.55 &  0.98&0.05 \\
			\bottomrule
		\end{tabular}
	
	\caption{We simulated from the bivariate probit with 25,000 observations and deploy our methodology.  cor refers to the correlation between predicted and true for the average causal risk difference (ACRD), and the rmse is the root mean square error. Fit using the `GJRM' package of \cite{marra}.}
	\label{bivartable_treat}
\end{table}

\begin{figure}[!httb]
	\centering
	\begin{subfigure}{.5\textwidth}
		\includegraphics[width=6.cm]{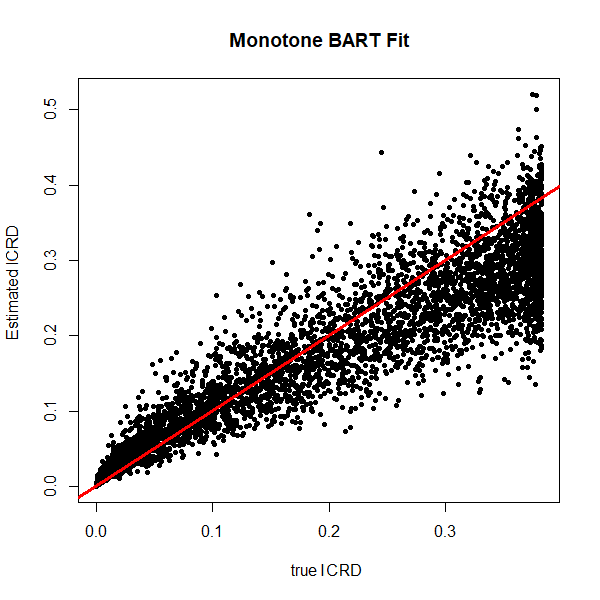}
	\end{subfigure}%
	\begin{subfigure}{.5\textwidth}
		\includegraphics[width=6.cm]{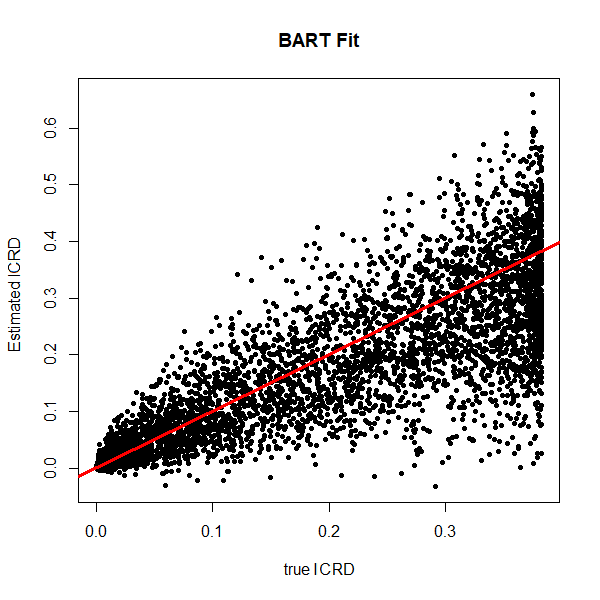}
	\end{subfigure}
	\caption{Plots of expected individual causal risk differences (ICRD) vs. our estimates, i.e. a plot comparing the difference in potential outcomes from the bivariate probit model ($\Phi(\alpha_0+\alpha_1\bm{x}_i+\gamma)-\Phi(\alpha_0+\alpha_1\bm{x}_i)$) versus our estimate.  In the DGP, $\rho=0.25,\gamma=1$.  The monotone BART correlation between $\tau$ and $\hat{\tau}$ is 0.928 and for BART is 0.826}
	\label{monovsnorm_ITE}
\end{figure}

It was stressed in the main document how using  a BART model with a monotonicity constraint improves our estimation of the ICRR, but the improvement is even more pronounced when studying risk differences.  
In figure \autoref{monovsnorm_ITE}, we look at the comparison of ICRD (individual causal risk difference) estimates from data generated by the bivariate probit, with the left hand side of our system of equation probabilites estimated with BART and monotone BART.  We display in the main file, but with inducements as the estimand of interest.

\section{Comparing machine learning methods for the observational data}\label{machine_append}
\begin{figure}[!t]
	
	\begin{minipage}[t]{.5\textwidth}

		\includegraphics[width=8cm]{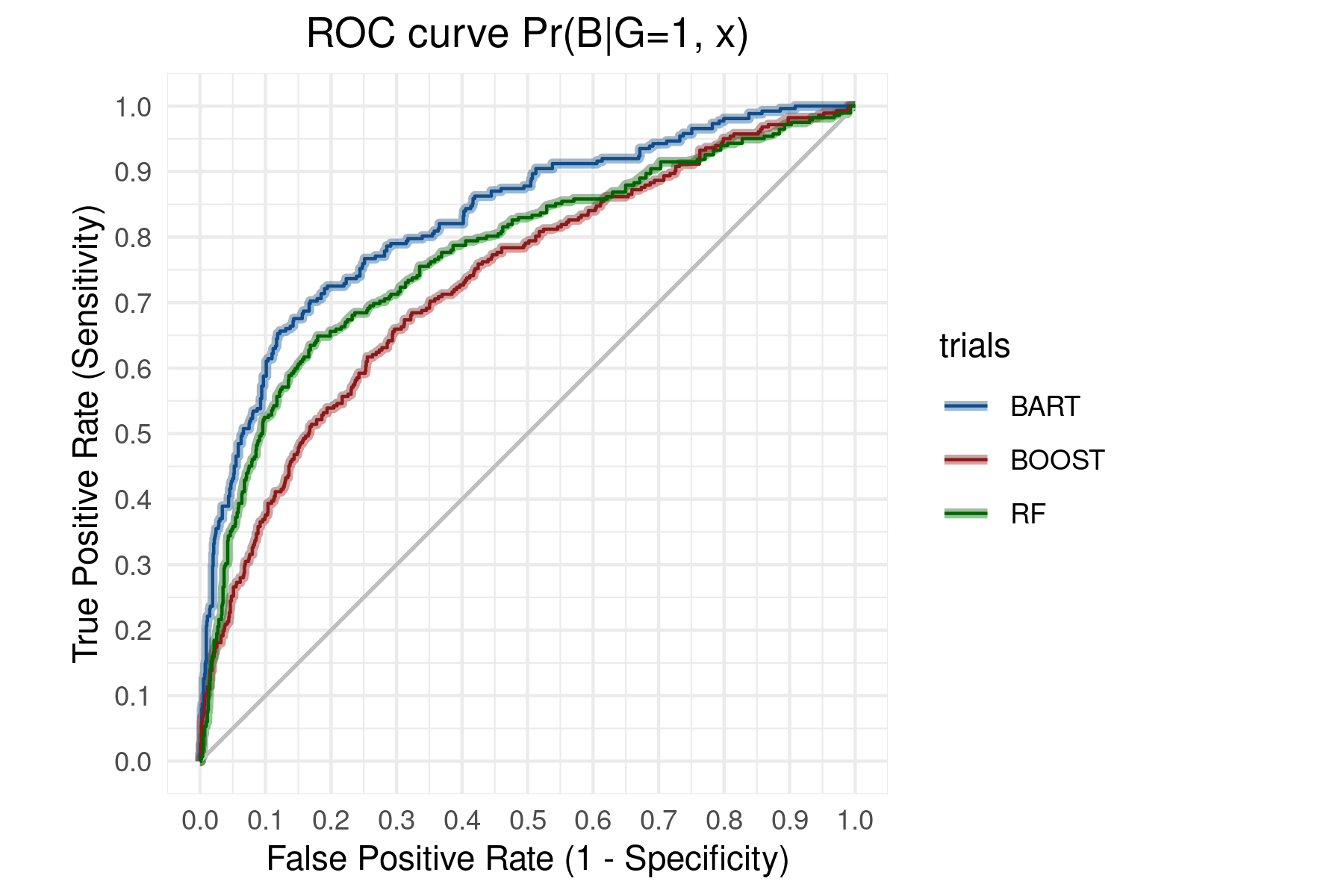}

	\end{minipage}\hfill
	\begin{minipage}[t][-2.2cm][b]{.5\textwidth}	
		\centering
		\begin{tabular}{P{1.1cm}P{1.2cm}P{1.25cm}P{1.25cm}}
			&\color{BrickRed}\textbf{XGBoost}&\color{darkgreen}\textbf{RanFor}&\color{navy}\textbf{BART}\\ \midrule
			
			Case   & AUC  & AUC & AUC  \\ \midrule
			$B_1, G_0$ &0.80&0.83&0.87\\ 
			\rowcolor{navy!49!white}$B_1, G_1$ &0.73 &0.78&0.83 \\
			$G_1$ &0.88&0.90&0.90 \\ \bottomrule
		\end{tabular}

	\end{minipage}
	\caption{Left: Area under curve of ROC plot, balanced 5-fold CV.  Plot of ROC performance for predicting $\Pr(B\mid G=1, \bm{x})$ .  Corresponds to shaded row in table on right.  Right:  case 1 is predicting bankruptcy when no going concern is issued, case 2 is when no concern issued, and case 3 predicts if concern is issued.  All the methods are \emph{similar}, with monotone BART seeming to be the top performer.}
	\label{roc_plot}
\end{figure}

Here we present our results from fitting the left hand side of equation(15) in the main file.
In this section, we compare the performance in predicting the left hand side of equation(15) using various non-parametric ``machine learning'' tools.  In particular, we compare using monotone BART \footnote{We use a BART \citep{bart} model for the $\Pr(G\mid \bm{x})$ scenario, consistent with the main text.}, random forests \citep{rf}, and xgboost \citep{boost}. Referencing \autoref{roc_plot} seems to indicate our methodology outperforms competitors in a cross-validation assessment\footnote{In our main text, we do not do a cross validation to obtain our probabilities, but rather get the probabilities from deploying the monotone BART models on the entire dataset.  In this case, our $B_1,G_0$ auc was 0.88, $B_1, G_1$ was 0.83, and $G_1$ was 0.92.}.

\end{document}